\pdfoutput=1
\documentclass{sig-alternate}
\usepackage{comment}
\usepackage{array}
\usepackage{url}
\usepackage{subfig}
\usepackage{listings}
\usepackage{algorithm}
\usepackage{algorithmic}
\usepackage{rotating}

\newcommand{\figref}[1]{Figure~\ref{#1}}
\newcommand{\secref}[1]{Section~\ref{#1}}

\newcommand{\tabref}[1]{Table~\ref{#1}}

\usepackage[font=small,labelfont=bf,textfont=bf]{caption}
\usepackage{xcolor}
\usepackage{colortbl}
\usepackage{xfrac}
\title{Data Driven Energy Efficiency in Buildings}

\author{Nipun Batra$^1$, Amarjeet Singh$^1$, Pushpendra Singh$^1$, Haimonti Dutta$^2$\\ Venkatesh Sarangan$^3$, Mani Srivastava$^4$\\ 
\small$^1$Indraprastha Institute of Information Technology Delhi, India ~\{nipunb,~amarjeet,~pushpendra\}@iiitd.ac.in\\
\small$^2$ CCLS Columbia ~\{haimonti@ccls.columbia.edu\}\\
\small$^3$ Innovation Labs, Tata Consultancy Services ~\{venkatesh.sarangan\}@tcs.com\\
\small$^4$ UCLA ~\{mbs@ucla.edu\}\\
}

\begin{document}

\maketitle

\begin{abstract}
\noindent
Buildings across the world contribute significantly to the overall energy consumption and are thus stakeholders in grid operations. Towards the development of a smart grid, utilities and governments across the world are encouraging smart meter deployments. High resolution (often at every 15 minutes) data from these smart meters can be used to understand and optimize energy consumptions in buildings. In addition to smart meters, buildings are also increasingly managed with Building Management Systems (BMS) which control different sub-systems such as lighting and heating, ventilation, and air conditioning (HVAC). With the advent of these smart meters, increased usage of BMS and easy availability and widespread installation of ambient sensors, there is a deluge of building energy data. This data has been leveraged for a variety of applications such as demand response, appliance fault detection and optimizing HVAC schedules. Beyond the traditional use of such data sets, they can be put to effective use towards making buildings smarter and hence driving every possible bit of energy efficiency. Effective use of this data entails several critical areas from sensing to decision making and participatory involvement of occupants. Picking from wide literature in building energy efficiency, we identify five crust areas (also referred to as 5 Is) for realizing data driven energy efficiency in buildings : i) instrument optimally; ii) interconnect sub-systems; iii) inferred decision making; iv) involve occupants and v) intelligent operations. We classify prior work as per these 5 Is and discuss challenges, opportunities and applications across them. Building upon these 5 Is we discuss a well studied problem in building energy efficiency - non-intrusive load monitoring (NILM) and how research in this area spans across the 5 Is.

\end{abstract}

% % Now added inline
% % \input{introduction}
% %
\section{Introduction}
\label{sec:introduction}
\noindent 

\begin{figure}
  \centering
  \includegraphics[width=\columnwidth]{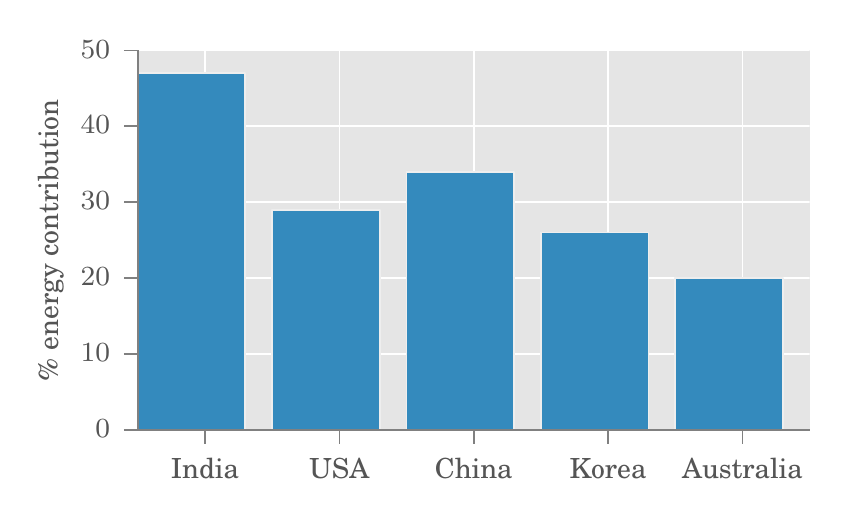}
  \caption{Contribution of buildings to overall energy consumption}
  \label{fig:country_proportion}
\end{figure}

\noindent Buildings across the world contribute significantly to total energy usage. \figref{fig:country_proportion} shows the contribution of buildings to energy consumption in India, USA, China, Korea and Australia~\cite{country_india, country_us, country_china, country_korea,country_australia}. Rapid rate construction of buildings presses the need to look into improving energy efficiency in buildings with a goal of decreasing overall energy footprint. Thus, towards the vision of sustainability in the age of dwindling natural resources, buildings need to be made more energy efficient.

``Measure twice and cut once"- so goes the old adage. With this adage in mind and the goal of understanding building energy efficiency, collecting building energy data assumes prime importance. Traditionally, building energy data included \textbf{monthly} electricity bills collected \textbf{manually} by the utility companies and other \textbf{sporadic} data such as energy audits. Owing to the manual and sporadic nature of this collected data, the available data is very sparse and provides limited insights into building energy efficiency.

Buoyed by the success of data sets such as MNIST~\cite{mnist} in instigating machine vision research, previous study~\cite{redd} suggests that research in building energy domain can be spurred by availability of data sets. While previously deemed improbable, collection of such data sets has become increasingly common due to low cost sensing devices enabling \textbf{high resolution} and \textbf{automated} data collection. Furthermore, governments and utilities across the world have started rolling out smart meters\footnote{UK:~\url{https://www.gov.uk/government/policies/helping-households-to-cut-their-energy-bills/supporting-pages/smart-meters}; \newline Australia:~\url{http://uemg.com.au/customers/your-electricity/smart-meter-rollout.aspx}} with an aim of developing a smart grid. Beyond the envisioned applications by the government and the utilities, this data can also be used for developing an understanding into building energy. Smart meters typically report electricity consumption to the utility and to the end consumer at rates ranging from a reading every second to a reading every hour. Commercial buildings are also increasingly managed with Building Management Systems (BMS) which control different sub-systems such as lighting and HVAC. BMS are computer based control systems for controlling and monitoring various building systems such as HVAC and lighting and are typically used to manage commercial buildings. BMS sense several spatially distributed points across the building to monitor parameters required for control action. Buildings are also commonly equipped with ambient sensors for monitoring parameters such as light, temperature and humidity. These ambient sensors are often coupled with security systems to raise intruder alarm or to maintain healthy ambient conditions via thermostat or BMS control. With the advent of smart meters, increased usage of BMS and ease of availability and installation of ambient sensors, there is a now a deluge of building energy data.

While collecting building energy data is easier than before, deployments can get increasingly hard to manage, especially when large number of sensors are deployed. However, previous research~\cite{granger,iawe,copolan} has highlighted that building energy data can be obtained at different spatio-temporal granularities. This highlights the need for \textbf{optimal instrumentation} that can have different connotations for spatial and temporal domain. For spatial domain, optimal instrumentation involves monitoring at a subset of locations while still being able to accurately predict at all the desired locations. For the temporal dimension, optimal instrumentation involves sampling at a lower resolution while still being able to predict at a higher temporal resolution. Desired application may choose the optimal set of sensors considering cost-accuracy tradeoffs across these different granularities.

%Building energy applications may employ different permutation and combinations of sensors to sense the desired phenomena. For instance, appliance level electrical usage can be found by instrumenting each individual appliance or by instrumenting only the mains and applying machine learning to estimate appliance usage. Thus, one needs to consider various facets such as desired accuracy and instrumentation costs in deciding how to \textbf{instrument optimally}.

Traditionally, there exist multiple building subsystems such as security, networking, HVAC and lighting, each performing their own operations in isolation. However, buildings are a unified ecosystem and optimal operations would require \textbf{interconnecting} these \textbf{sub-systems}. The combined information from the different sub-systems is greater than sum of individual information from each of the systems. 

%Traditionally networking equipment such as routers and access points are used only to manage data communication and meeting software and chat clients are used for scheduling meeting. However, 
%Thana\-yankizil et al. ~\cite{softgreen} fuse information from networking equipment with Office chat and meeting software for detecting occupancy for efficient HVAC control.
\begin{figure}
  \centering
  \includegraphics[scale=0.3]{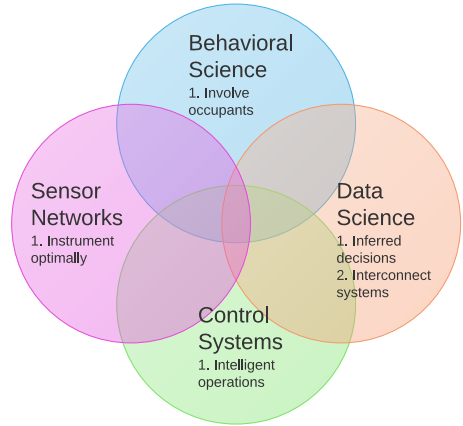}
  \caption{Data driven building energy efficiency sits at the intersection of behavioral science, data science, control systems and sensor networks}
  \label{fig:building_energy}
\end{figure}

%While each data source in the building may provide rich information, richer information can be obtained if we \textbf{interconnect systems} and data sources. For instance, if we detect water usage and stove usage at the same time, we can infer that occupants may be cooking. On similar lines, Thanayankizil et al. ~\cite{softgreen} fuse information from networking equipment such as Wi-Fi access points with Office chat and meeting software for detecting occupancy. By interconnecting lighting system and motion sensors, one can obtain presence based lighting control.

Once the systems are interconnected, data coming from diverse systems can be used to improve upon the decision making for optimal building operations. \textbf{Inferred decision making} can help in identifying inefficiencies, raising alerts and suggest optimizations. Traditionally data from within the system has been used for simple decision making e.g. motion sensor based lighting control.
%, PLC based HVAC control based on return air temperature and return humidity. 
Furthermore, utility companies previously relied on customers' phone calls to detect power outages. However, utility companies can now leverage smart meter data from different homes to quickly detect power outages and plan accordingly. 
%Studying the correlation between HVAC energy usage and ambient temperature can be used to determine optimal set point temperatures within a commercial building. 
Temporal patterns in the electricity consumption data~\cite{energy_dashboard,iawe, hart_1992} can be used to detect faulty operations, predict load profiles and optimize operations accordingly. Spatial patterns in the electricity consumption can identify different electrical sub-systems and study their effect on aggregate consumption. The vast set of analysis possible further necessitates the importance of data sets collected from the real world which can be to used to simulate the effect of inferred decisions before performing optimizations.

%This is where you may also mention that for building such sophisticated analysis, it is useful if real datasets are used to validate their applicability which in turn shows the dependence on optimal sensing.
%The rich data collected by optimal instrumentation and system interconnection can be leveraged for \textbf{inferred decision making}. 
%Rules of thumb and inefficient mechanisms were previously used for deducing common building applications. 

% % COMMENTED AS NOT REALLY IMPT % %
%The inferred decisions can be translated into control actions culminating in optimizing energy efficiency and reducing costs via the following two means: i) by involving occupants and ii) by automated control.

Buildings constitute ecosystems involving interactions between the occupants, the physical and the cyber world~\cite{hbci,satyanarayanan}. The presence of occupants, each with their individual preferences makes the ecosystem even more complex. Traditionally, the control decisions for building operation are taken at a central facility level assuming certain desired operating conditions without involving the occupants in decisions regarding desired conditions. Such policies are bound to be energy inefficient and may cause occupant discomfort at times~\cite{batra_issnip}. Moreover, previous literature~\cite{energy_buildings_survey} suggests that energy unaware occupant behavior can add upto one-third to a building's energy performance. However, when empowered with actionable feedback, occupants may save upto 15\% energy~\cite{darby_2006}. Building occupants can also provide useful data such as comfortable temperature and light intensity levels, which can be used to optimally schedule the HVAC systems. Thus, various optimizations in the building energy ecosystem can be enabled by \textbf{involving occupants}.

All of optimal sensing from interconnected subsystems, resulting in decisions inferred from the rich dataset while involving the occupants will overall result in \textbf{intelligent operations}. Such intelligent operations already exist but miss out one or more aspects discussed previously i.e. either they do not involve the occupants or are taken at a subsystem level without accounting for data from other subsystems in operation. 

Thus, based on literature in building energy domain, we have identified these five crust areas also called five Is which are as follows: i) instrument optimally; ii) interconnect sub-systems; iii) inferred decision making; iv) involve occupants and v) intelligent operations. \figref{fig:building_energy} summarizes the relationship amongst the 5 Is of building energy which span across broad fields such as sensor networks, behavioral sciences, data science and control science. The interconnected nature of these 5 Is makes the overall optimization of building energy efficiency a complex problem. For solving such a complex problem of intelligent building operations, rich data from across the diverse systems, together with smart algorithms for data inference and participatory engagement of occupants is critical. Such rich datasets are now becoming a reality~\cite{redd,blued,iawe,smart}. Further, rich research exists in from recent times that address algorithms for data inference and understanding behavioral aspects of occupants feedback~\cite{johnson_2013,darby_2006,bidgely_2013}. This work bring together these different aspects into a survey while pointing out the big opportunities that exist along each of the 5 Is together with those existing at the system level. 

\section{Example scenario}
\label{sec:scenario}
\noindent Having briefly discussed the various facets of data driven energy efficient buildings, we now discuss an example scenario covering all of these aspects. We base our theme around the classic paper on Pervasive computing~\cite{satyanarayanan}.

% Most of the pieces of this crossword exist in isolation and need to be joined together to complete the picture. 

John goes to bed at 11 PM in the night. He sets the alarm on his smartphone at 5 AM when he intends to go for a jog. However, his health has not been at the best since the last few days affecting his sleep. He decides to inform his alarm clock that it should wake him up at 5 AM if he is able to sleep by 11:30, else wake him up after he has completed 6 hours of his sleep. John's smart surround sensors capture his sleep patterns and communicate to the alarm to ring at 6 AM. His sleep disturbances, coughing are also captured, archived and emailed to John for sending to his doctor. John goes to the refrigerator to grab some cold water. The refrigerator understands that the person who has come to pick up water is John. John's heath system reminded his refrigerator that John has been advised to refrain from cold water. John gets ready and leaves home for office at 9 AM. The security system, door and motion sensor register this movement. This information is communicated to John's thermostat which starts ramping down. His lights also turn off automatically sensing his absence. 
%As John steps out, his phone is fed in with real time traffic and weather updates suggesting him the best route to reach his office. 
John has a meeting at 10 AM and thus he would only go to his cabinet to keep his bag. Based on the number of expected attendees, the thermostat in the meeting room starts ramping up, in order to achieve the desired temperature in time for the meeting. 
%Gradual ramping up is much more energy efficient.

John's washing machine at his home is scheduled to run for 1 hour before John returns in the evening. His washing machine interacts with the grid and predicts the best time to run when the load on the grid is low is noon time when the electricity prices are low. Around 3 PM, winds carry away the clouds and there is bright sunshine. John's solar system starts producing electricity. Part of this DC produce is directly fed to DC appliances in John's home and the surplus is stored in a battery. John figured that he can save more money by locally consuming his generated solar energy as opposed to selling it to the grid. In the meanwhile, some of the artificial lights in John's office turn off in lieu of the sunlight available. John gets an email from the NILM system installed at his home about his disaggregated monthly consumption. The system recommends that the tungsten based lighting in his home is very inefficient and eating up 30\% of his bill. If John would replace the same with more efficient LED based lighting, his overall bill would go down by 15\%. Considering the cost of replacement, his ROI period would be less than 6 months, after which he would be able to save 10 USD a month. 

We now analyse the above scenario and see how each small piece in this giant picture is a reality today. Systems such as iSleep~\cite{isleep} have explored using smartphones and motion sensors for sleep quantity and quality detection. In order to wake up John after he has completed 6 hours of sleep, his smartphone detects his sleep pattern. 
%Possibly the smartphone application offloads some of the computation to a cloud based service were John based on the available battery levels. 
His smartphone contains an app which detects his cough patterns~\cite{coughsense} and uploads the most critical data to his doctor. When John wakes up and goes to the refrigerator, systems for energy apportionment~\cite{apportionment} ascertain that it is John and not someone else who is trying to draw cold water. The refrigerator is connected to John's smartphone over IP and informs him that cold water could be injurious to his health. Several years of research in occupancy detection using variety of sensing modalities~\cite{occupancy_ucsd,smart_thermostat,niom_eth,granger} are able to detect that John's home is unoccupied when John leaves for his office. Smarter occupancy driven thermostats have been proposed in recent literature~\cite{smart_thermostat,myjuolo} which control and save energy based on occupancy prediction and external temperature, when John leaves for his office. By interconnecting soft sensors~\cite{softgreen} such as office chat client and meeting software, the thermostat in John's personal cabinet does not ramp up since he has a meeting in the board room.
Conference room management sensor system~\cite{conference} in the boardroom is alert to the number of occupants and drives the HVAC accordingly.

Meanwhile in John's home his washing machine turns on at 1 PM. Since this is an off-peak period, electricity is available at much cheaper rates at this time in comparison to the rest of the day. John scheduled his washing machine to run for one hour when the electricity would be cheap,	 which was enabled by demand response strategy~\cite{smartcap}. Due to availability of sunlight after 3 PM, DC appliances in John's home switch from the utility and consume raw DC power from the solar panel~\cite{picogrid}. The lighting control system~\cite{lightwise} in John's office also senses the bright sunshine and saves power while maintaining comfortable lighting levels inside the office. The smart meter installed at John's home is regularly collecting his electricity data and periodically sends John the disaggregated breakdown of power by appliances~\cite{hart_1992}.

While many of these individual systems have been explored in the past in isolation and often in research settings, their application to the real world and their interconnection to complete the picture remains largely untested. The stepping stone to making such complex scenario a reality is the high resolution multimodal data collected from each of these subsystems, interconnected  together, with algorithms that can do efficient operations while accounting for John's preferences.

% % \input{building_energy_data}

\section{Building Energy Data}
\label{sec:building_energy_deluge}
\begin{figure*}
  \centering
  \subfloat[\scriptsize London, UK]{
  \label{fig:london}
  \includegraphics[scale=0.13]{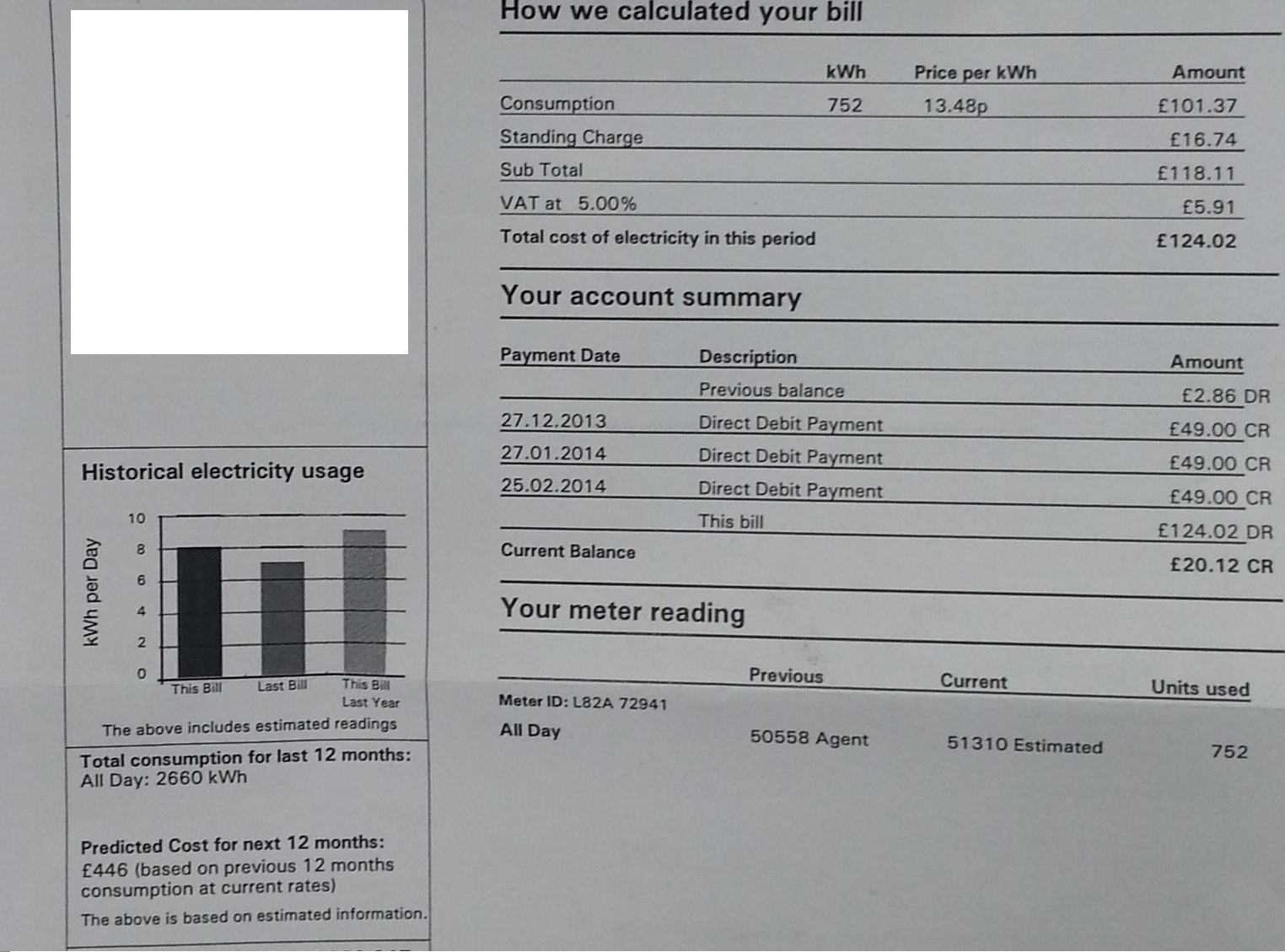}}
  \subfloat[\scriptsize Delhi, India]{
    \label{fig:delhi}
    \includegraphics[scale=0.08]{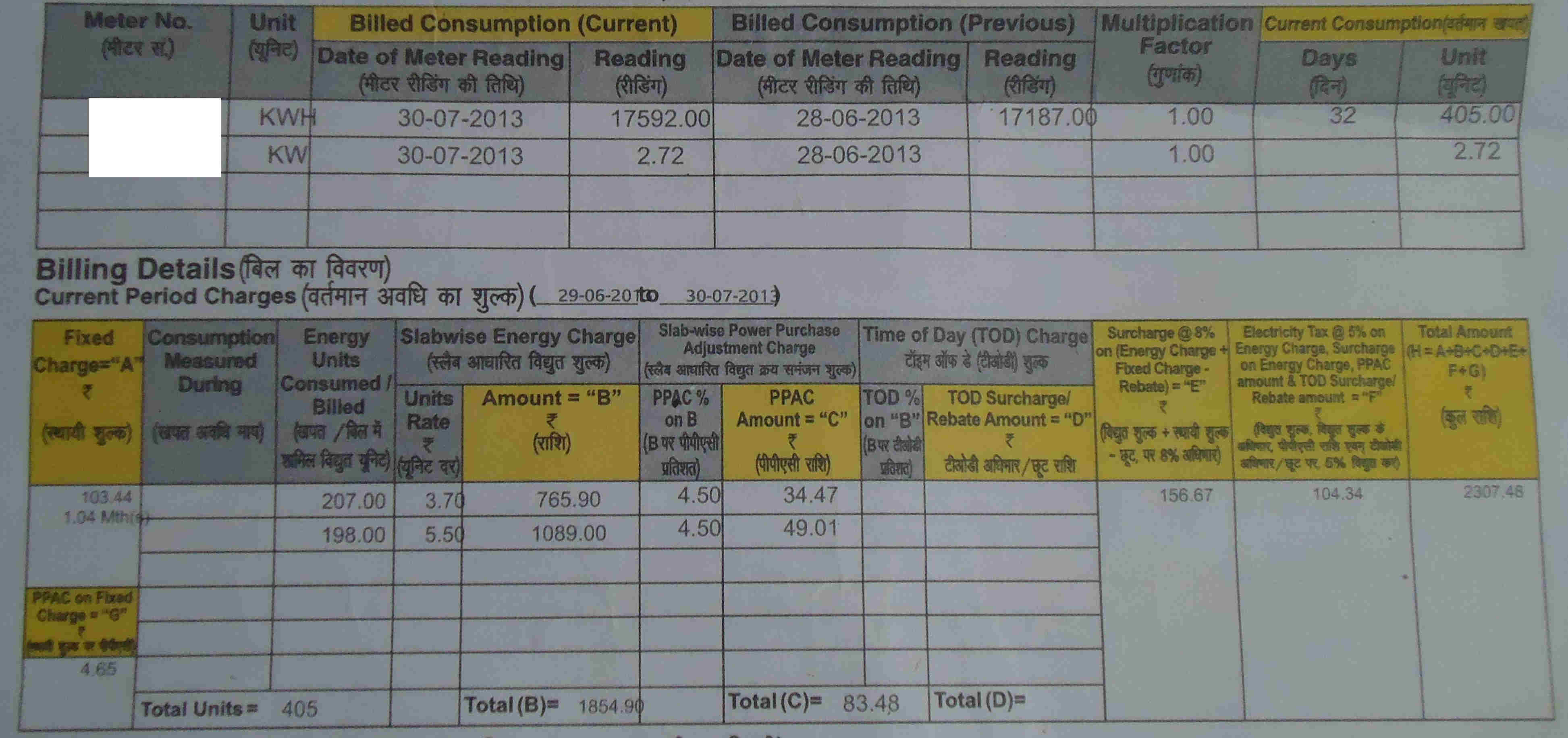}}
  \caption{Electricity bills from residential apartments in London, UK and Delhi, India. [Best viewed in color]}
  \label{fig:bill}
\end{figure*}

\begin{figure}
\centering
\includegraphics{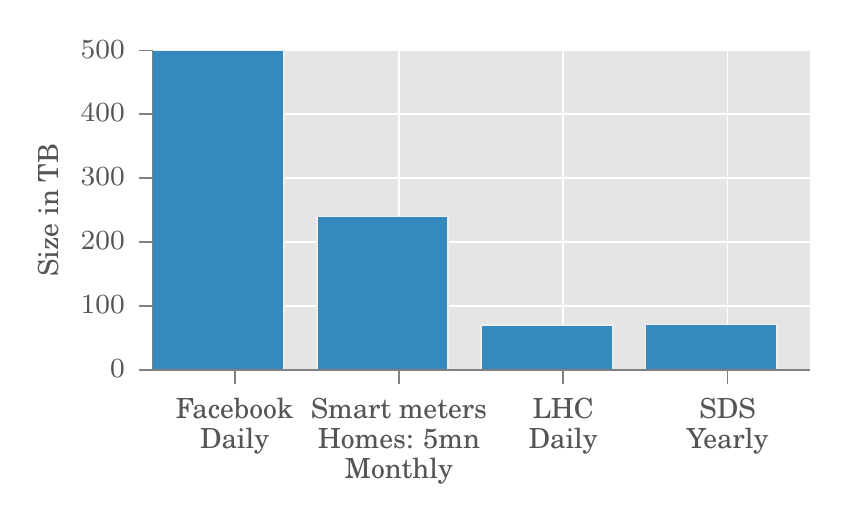}
\caption{Comparing the data deluge in building energy with social networking (Facebook), particle physics (Large Hadron Collider(LHC)) and astronomy (Sloan Digital Sky (SDS))}
\label{fig:big_data}
\end{figure}

\noindent ``Data is the new oil". Data science has brought about a paradigm shift in the way problems are solved in myriad applications. From social networks to astronomy and subatomic physics, data has brought a new revolution enabling answering important questions. We now briefly discuss how data can play a pivotal role towards the development of energy efficient building.

Traditionally, building energy data was collected only for billing purposes. \figref{fig:bill} shows the electricity bills of residential apartments in London, UK and Delhi, India. Usually, data for such bills is collected once a month manually by the utility companies. The bill shown in \figref{fig:delhi} only provides the units of energy consumed during the billing period. Commercial entities owing to their heavy consumption are often priced on a time of day based pricing. Thus, in addition to the total units consumed, commercial entities are also provided with units consumed in peak and non-peak hours. In both of these cases, the consumption information is coarse and provides limited actionable insights. 

\begin{figure}
  \centering
  \includegraphics{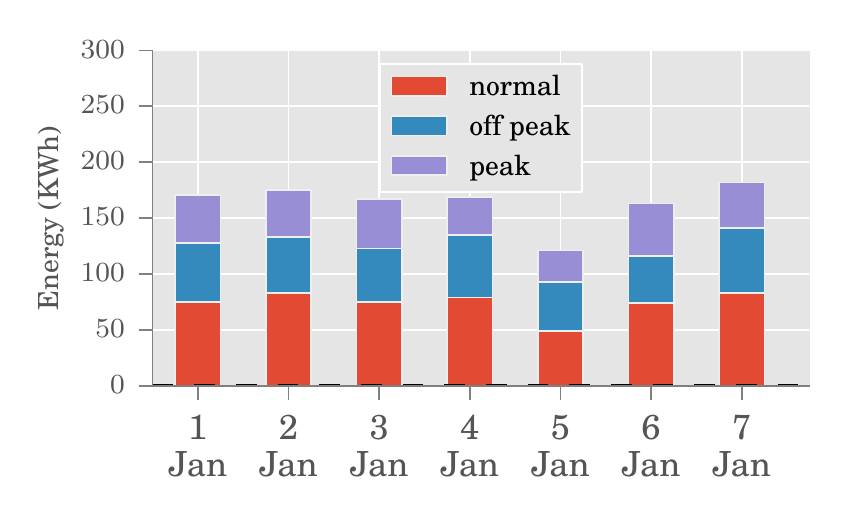}
  \caption{Time of day energy usage for a building in IIIT Delhi campus}
  \label{fig:bill_tod}
\end{figure}

\begin{figure}
  \centering
  \includegraphics{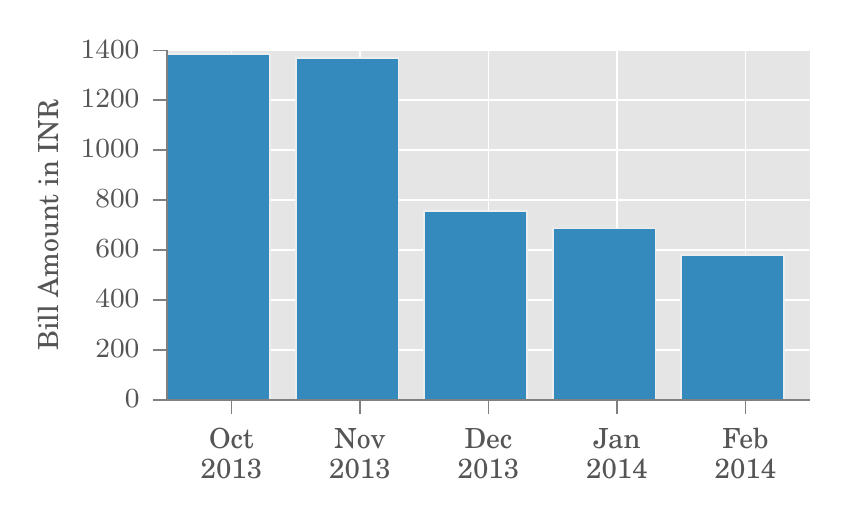}
  \caption{Historical bill collected from the home used in iAWE data set~\cite{iawe}}
  \label{fig:historical_bill}
\end{figure}

With the advent of the smart meter and some of the other sensors discussed in \secref{sec:introduction}, more building energy data is now available than ever before. While the conventional process followed by utilities would generate a single reading per home per month, smart meters collecting data once every 15 minutes would be collecting 3000 times more data~\cite{smart_grid_data_deluge}. Smart meters are often capable of collecting data at higher rates of once every minute which can amount to 240 TB of data collected from 5 million homes a month~\cite{autogrid_data_deluge}. \figref{fig:big_data} contrasts the volume of this smart meter data collected  with other applications which have greatly benefited by the data deluge. The availability of such building energy data enables one to answer several important questions, some of which are presented below: 
\begin{itemize}
\item Consumers can be informed in real time about their consumption. This may be seen analogous to the cell phone alerts we get after every data transaction or call made. This may greatly help the end users to not only understand their usage but also act on it to conserve energy.
\item Utilities earlier used to rely on receiving phone calls from aggrieved customers for detecting outages. However, with smart meter data readily available, utilities can leverage it for speedier outage detection and allocate resources efficiently to reduce downtime.
\item \figref{fig:bill_tod} shows the time of day energy consumption of a building in IIIT Delhi campus. When such detailed information is made available to the end users, they can take initiatives towards shifting their loads to normal or off-peak hours, when electricity is cheaper. Such measures also benefit the utility as the peak demand is reduced. 
\item \figref{fig:historical_bill} shows the trend in energy consumption from a home in New Delhi~\cite{iawe}. This decreasing energy usage can be attributed to decreasing temperatures and hence decreasing use of air conditioners. Developing an understanding into the correlations between weather and energy usage can allow better HVAC control. Such information is present in the UK bill (\figref{fig:london}). However, without detailed analysis the same does not translate into actionable savings.
\end{itemize}

Towards the realization of many of the above mentioned questions, building energy data requirements need to be categorized as per the application scenario. We firstly discuss the different temporal and spatial resolutions at which such data can be collected.

\noindent \textbf{Temporal resolution:} In accordance with the intended application, building energy data is collected at different temporal resolutions. Energy audits are performed once every few years~\cite{energy_auditing} and provide key insights into several aspects of buildings including energy efficiency and comfort levels. Such audits are costly and require significant instrumentation. At a lower resolution of once a month utility companies collect total energy consumption. As discussed previously this provides limited insights to the end user. Commercial entities often monitor their power factor daily as they are liable to be penalized if the power factor drops below a certain threshold. Most of the above considered data collection is largely manual and sporadic. Monitoring at higher resolutions of 1 hour or lesser requires automated mechanisms such as installation of smart meters. With the national smart meter roll outs across many countries as discussed in \secref{sec:introduction} and initiatives such as Greenbutton\footnote{\url{http://www.greenbuttondata.org/}}, collecting and accessing electricity data has become much easier for the end consumers. Certain applications such as non-intrusive load monitoring (NILM)~\cite{hart_1992} may require high frequency data with sampling rates of more than a thousand samples every second~\cite{electrisense,blued}. We summarize the temporal variations in building energy data collection in \tabref{tab:temporal}.

\begin{table}
\centering
\scriptsize
\begin{tabular}{cc}
\hline
\textbf{Application} & \textbf{Rate}\\ \hline
Energy auditing & once every few years\\
Electricity billing & once a month\\
Power factor checking & once a day\\
Automated meter reading & once 15 minutes or less\\
High frequency NILM & several KHz\\ \hline

\end{tabular}
\caption{Temporal variations in building energy data collection}
\label{tab:temporal}
\end{table}

\begin{figure}
 \centering
  \includegraphics[scale=0.3]{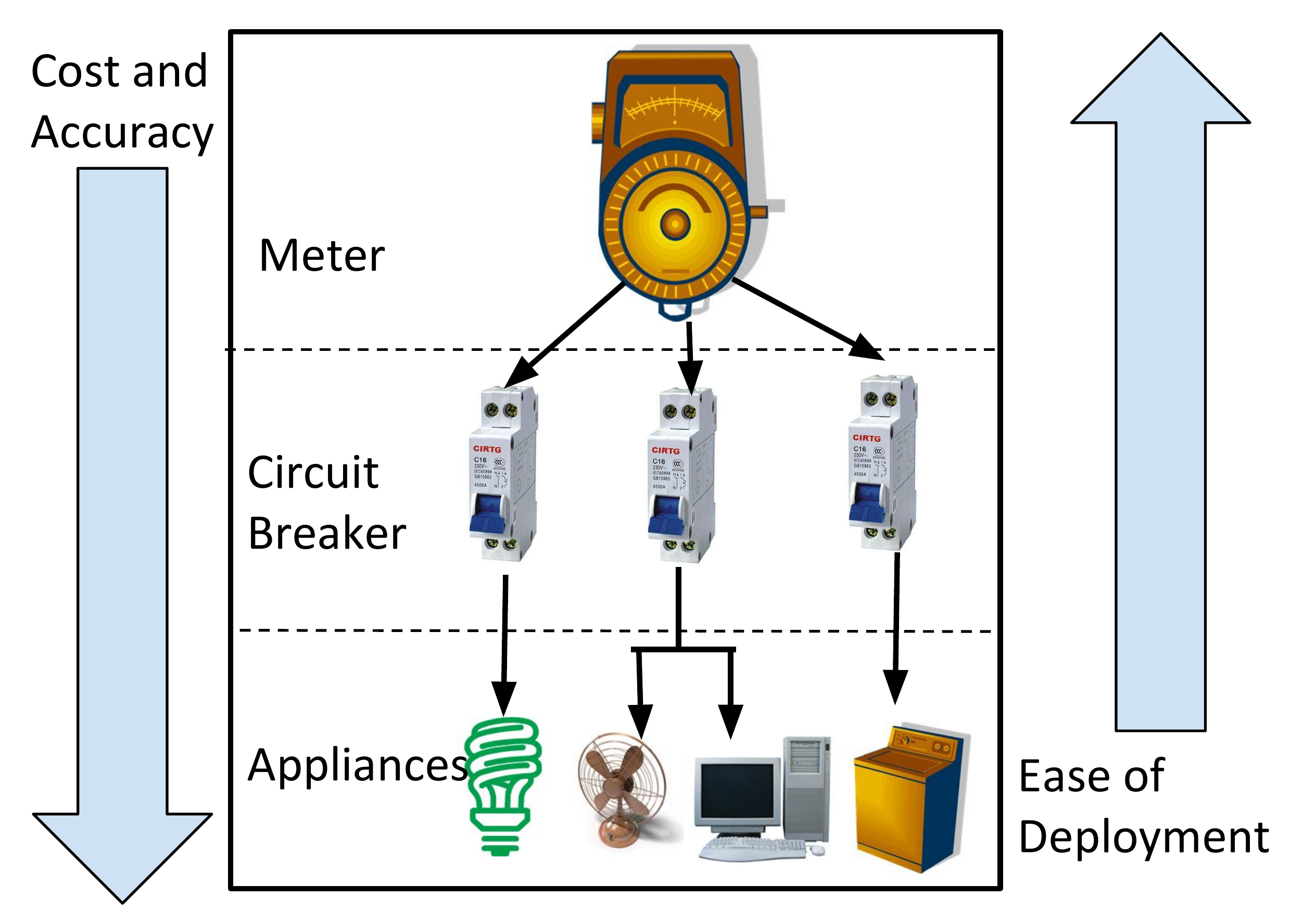}
  \caption{Cost-accuracy optimization in obtaining building appliance level data}
  \label{fig:optimal_instrumentation}
\end{figure}

\noindent \textbf{Spatial resolution:} Buildings consist of multiple location contexts such as floors, rooms, wings. The utility (water, electricity, gas) networking in buildings also follow different topologies introducing the spatial context across different nodes in the flow. Thus, different location contexts and utility topologies provide different spatial resolution to observe building energy and related phenomena. The data collected at different spatial resolutions can provide a varying degree of information~\cite{spatial}. For instance, an electric meter connected at the supply from the grid provides a single point of easy to gather data. This data is also used by the utilities for billing information. For finer information, electric meters can be installed at sub-system level (e.g. HVAC, Genset in commercial buildings and circuits at residential buildings) or at appliance level. As we traverse down the electrical load tree (\figref{fig:optimal_instrumentation}) in a building, i.e. supply from grid to subsystems such as HVAC to individual appliances such as printers, information content is enhanced. However, this additional information comes at the cost of increased monetary and time cost of sensing and maintaining additional instrumentation. 

Having discussed the spatio-temporal categorization of building energy data and its sources, we now discuss different sensing modalities which exist within buildings and can provide additional insights alongside energy data. 
\begin{enumerate}
\item \textbf{Building management systems (BMS):} Using open standards such as BACNet, the data collected by the BMS can be used for building energy analysis. BMS systems can provide rich data for building energy applications and can also be used for control applications based on proposed optimizations. Typically, BMS control HVAC, lighting and security systems and may use a variety of sensors such as:
\begin{itemize}
 \item cameras, motion, RFID for security and access control
 \item temperature for HVAC monitoring
 \item light for lighting control
 \item motion and RFID for occupancy monitoring
\end{itemize}
 
\item \textbf{Water meters:} Several building sub-systems comprise of a rich interaction between electricity and water. For instance, the central HVAC systems involve heating or chilling water to heat or cool the air respectively. Further, water treatment plants have a significant energy footprint. In several buildings, electrical pumps are often used to store the water in tanks for times when water is not available. Swimming pools and water sprinklers also sit at the intersection of energy and water. Thus, water meter information measuring parameters such as flow and pressure may provide significant hints in understanding energy usage of such building sub-systems.
\item \textbf{Ambient sensors:} Ambient conditions significantly impact the working environment~\cite{energy_buildings_survey}. Hence, many of ambient parameters such as humidity, light level, temperature are monitored for maintaining good working conditions. While some of the monitoring equipment are controlled centrally by the BMS, others such as smart thermostats can exist as independent units. 
\item \textbf{Access and security systems:} These systems log entries into rooms and can be used to estimate the occupancy levels in near real-time. 
\end{enumerate}

Having discussed the data deluge and its categorization, we move towards instrumentation systems for obtaining such data.

% % \input{instrument}
\section{Instrument Optimally}
\label{sec:instrument}
\noindent Having discussed the importance of data in realizing building energy efficiency, we discuss how to efficiently obtain such data. Firstly, we discuss general sensor deployments followed by discussion in the specific context of building energy as per the spatio-temporal categorizations which we discussed previously in \secref{sec:building_energy_deluge}.

Sensor deployments have been well studied in the literature and are understood as means of providing data for improved operations across a variety of application domains~\cite{sn_survey_1}. Based on previous work in sensor network domain~\cite{hitchhikers_wsn, scale,sn_survey_1}, we summarize the design goals of such instrumentation as follows:
\begin{itemize}
\item \textbf{Low power consumption:} It is desired that the deployed sensors and/or controllers do not consume significant power impacting their life (when powered from a battery) or utility (when powered from mains). 
\item \textbf{Wide network coverage:} For collecting spatially distributed information, more than one end controller or node may be deployed. Often, these end nodes relay the collected data to a central node or gateway for storage or further processing. The deployment must ensure that all nodes are within the communication range of at least one node and together cover the whole area that is to be monitored, under different real world scenarios. Often, multi-hop strategies are used to increase network coverage. 
\item \textbf{Robust:} The real world presents many unforeseen challenges which the sensor deployment must account for. It is often desired that the system is capable of easy healing and recovery.
\item \textbf{Ease of deployment and maintenance:} Ease of deployment and maintenance outside the controlled settings has always been an important challenge and design goal of sensor network deployments. 
\end{itemize}
	
Sensor deployments for building energy monitoring involve some key peculiar features when compared to sensor deployments in general. Many of these features have made collecting building energy data a non-trivial task. Previous literature~\cite{hitchhikers_residential,iawe,batra_issnip} identifies these key differences, some of which we highlight as follows:
\begin{itemize}
\item \textbf{Ugly retrofitting:} While monitoring energy consumption is important, the occupants are also concerned about the building aesthetics.
\item \textbf{Hostility of building environments:} In general, building environments are less hostile when compared to external world deployments. However, accidental wire snapping and sensor mishandling are common.
\item \textbf{Buildings may become inaccessible after installation:} Since people work inside office buildings and stay in residential buildings, buildings often become inaccessible as the occupants resume normal life once the installation phase is complete. This makes debugging and iterative research difficult in the context of buildings.
\item \textbf{Occupant interaction drops with time:} After the initial interest, occupants usually get busy in their routine lives and forget to provide regular feedback pertinent to the deployment.
\item \textbf{Wireless spectrum may be clogged:} Many user appliances and gadgets operate in the same frequency as some of the deployed sensors. This is due to the limited availability of license-free frequency bands\footnote{Information about license-free bands in India:~\url{http://www.wpc.dot.gov.in/faq.asp
}} (which also vary across different countries). Limited frequency bands results in most sensors being designed to communicate in 2.4 GHz band, which is also used in household WiFi routers. This can cause interference and reduced user experience for the occupants and sensor data loss~\cite{iawe}.
\end{itemize}

Sensor deployment architecture for building energy monitoring typically involve end sensor nodes communicating with a central controller. This controller is primarily responsible for data aggregation and sensor node control. We now discuss the communication and automation protocol standards used for monitoring and control in building energy domain.

\subsection{Communication and automation protocols}
\noindent Over the past few years, many specialized interfaces for communication and automation have been standardized for building energy monitoring and control. Firstly, we discuss about building automation protocols. Building Automation and Control Network (BACnet)\footnote{\url{http://bacnet.org}} is a non-proprietary protocol used for automating and controlling HVAC, lighting, security and several other building systems. BACnet fundamentally consists of collection of objects called properties (an object is an information units such as temperature reading) and services for device discovery and accessing and setting these properties. It provides various data and physical links such as Ethernet and RS-232~\cite{building_automation_protocols}. The Modbus protocol developed by Modicon Inc. in the 1970s is widely used for building automation. Like BACnet, Modbus is also an open standard. Modbus provides client-server communication via its messaging structure. Recently, Modbus/TCP has been developed allowing Modbus communication over TCP/IP networks, in addition to allowing for serial communication~\cite{building_automation_protocols}. Similarly, LonWork and LonTalk are proprietary protocols developed by Echelon corporation. Extensive comparison of these communication standards has been well studied in the literature~\cite{building_automation,building_automation_1,building_automation_protocols}.

The term home automation is often used in the context of residential buildings to discuss building automation. Residential buildings usually have simpler electric wiring and smaller physical area in comparison to the commercial buildings. Due to these two reasons, several home automation protocols exploit existing power line infrastructure which work well over small distance electrical circuit (such as in homes), but may work poorly in commercial buildings. Powerline protocols such as Insteon and X10 have been widely used for home automation despite the low bandwidth and packet collision. However, these protocols were not designed for high frequency monitoring and need to be extended. Irwin et al.~\cite{insteon_ha} work around these shortcomings by creating smart polling strategies to maximize energy measured with least packet drop in their energy monitoring system based on Insteon.

In the other end of the spectrum, several protocols operating on wireless nodes are commonly used. Zigbee\footnote{\url{https://www.zigbee.org/Specifications/ZigBee/Overview.aspx}} and its close variant 802.15.4 present networking stacks similar to the layered internet protocol suite. Zigbee is intended to be low power and some of its key features include AES-128 security, multiple star topologies and personal area networks (PAN) and device discovery among others. Similarly, ZWave supports mesh topology and is commonly used in home automation. 

The internet of things (IoT) revolution has led to a directed effort on IP based sensor data monitoring. This allows physical sensors to be exposed as HTTP services allowing an integration with the ubiquitous internet and simultaneously allowing modern web standards based rich applications. The IoT revolution has also led to the popularity of constrained application protocol (CoAP) and IPv6 over Low power Wireless Personal Area Networks(6LoWPAN). CoAP implements the HTTP REST scheme without much of HTTP's complexities and supports only datagrams in the transport layer. 6LoWPAN defines header and encapsulation format for communication with 802.15.4 networks. \tabref{tab:home_automation} summarizes comparison of different home automation protocols.

\begin{table}
\centering 
\scriptsize
\begin{tabular}{p{1.1cm}p{1.4cm}p{1.1cm}p{1.1cm}p{1.1cm}}
\hline
\textbf{Protocol}&\textbf{RF band (MHz)} & \textbf{Range (m)} & \textbf{Bit rate(kb/s)} & \textbf{Routing}\\ \hline
ZigBee&868/915/2400&10-100&20/40/250&mesh, tree, source\\ \\
6LoWPAN&868/915/2400&10-100&20/40/250&RPL\\ \\
ZWave&868/908/2400&30-100&9.6/40/200&source\\ \\
Insteon&904&45&38.4&simulcast\\ \\
X10&310/433&100&20&simulcast\\ \hline
\end{tabular}
\caption{Comparison of home automation protocols. Mostly referred from~\cite{home_automation_survey}}
\label{tab:home_automation}
\end{table}

\subsection{Criteria for optimality}
\noindent Having discussed the important differences in building energy domain from the generalized sensor network domain, we discuss the issue of optimal instrumentation and how these differences are to be incorporated. We pick up electricity monitoring in a residential home to discuss the criteria for optimal instrumentation. As discussed previously in \secref{sec:building_energy_deluge}, electrical wiring in buildings create a natural hierarchy across different spatial resolutions. \figref{fig:optimal_instrumentation} shows these different spatial levels at which building energy data for a residential home may be obtained. As we go from the meter to the circuits and to the appliances, cost of instrumentation and maintenance increases. However, going this route from the meter to the appliances also adds significant information at each level. The optimal spatial resolution for instrumentation must be decided in accordance to the intended application. 

For the same application of electricity monitoring, sensors with different sampling frequencies exist. As the sampling frequency increases, so does the cost of the instrumentation and consequently the cost of maintaining more data. For some applications such as energy dashboards~\cite{energy_dashboard}, low sample data might suffice, whereas for applications such as high frequency NILM analysis~\cite{electrisense}, one may need to sense more than a thousand samples every second.

%Communication interface is another optimality criterion. Depending on the requirement of data throughput and whether an infrastructured or infrastructureless deployment is needed, one may decide amongst the communication and automation interfaces.
% discussed.
 
\subsection{Challenges and Opportunities}
Besides being peculiar in certain ways, building energy sensing also brings in a peculiar set of challenges and opportunities, which we discuss as follows:
\begin{itemize}
\item \textbf{Indirect sensing:} Owing to the challenges in maintaining and scaling sensor deployments, recent research~\cite{softgreen} proposed using existing sensor infrastructure such as Wi-Fi access point coupled with soft sensors such as office messenger for energy management of green buildings. Such research highlights the need to look into indirectly monitoring the desired parameter via the use of side channel information  or information from other existing context sources, which can help avoid new instrumentation. On the same lines, Kim et al. proposed ViridiScope~\cite{viridiscope}, a system which uses indirect sensing modalities such as magnetic, light and acoustic for power measurement of different appliances. Similarly, Schoofs et al. propose Annot~\cite{annot}, a system to automatically annotate electricity data using wireless sensor networks. Similarly, Barker et al.~\cite{smart} show the relationship between a refrigerator's power consumption and internal temperature. They key opportunity lies in understanding the causal relationship in data measured by different sensors to optimally sense a physical phenomena. However, this remains a challenging task owing to the per indirect sensor calibration required~\cite{viridiscope}.
\item \textbf{Remote configuration:} Buildings usually become physically inaccessible to researchers after the initial deployment phase of their study. As a result debugging becomes hard. Previous research~\cite{hitchhikers_wsn} has suggested that reconfiguration of sensor nodes should be preferred over reprogramming. Previous work~\cite{leap} has also mitigated this problem via remote debugging over IP. This challenge can be approached from various angles. On one front, human computer interaction (HCI) research can help identify mechanisms of allowing home occupants to easily configure their deployment (which is considered a highly technical task currently). From another angle, this problem has also been mitigated to a certain extent by using an IP backbone for all data communication. Another approach could involve distributed debugging and configuration often used in swarm based robotics.

\item \textbf{Unreliable network:} Previous studies~\cite{iawe, hitchhikers_residential} have highlighted unreliable network as one of the issues in building sensor deployments. Local buffering on the controller or the sensor (such as on device EEPROM) have been used as quick fix solutions to these issues. The opportunity here is to re-think about new architectures that can mitigate data loss due to unreliable network.
%\item \textbf{Privacy concerns:} Kim et al. ~\cite{challenges_residential} cite user privacy as an important challenge in resource monitoring for residential spaces. It remains an open challenge to find the sweet spot in data collection and preserving user privacy.
\item \textbf{Optimal sensor placement:} Finding the optimal number of sensor and their placement is a fundamental Wireless Sensor Network (WSN) task. While we took an application driven discussion on optimality criterion earlier, this problem can also be studied theoretically. This problem may be viewed as a spatial optimization problem over the number of sensors, communication costs and sensor information. Krause et al.~\cite{near_optimal} presented a Gaussian Processes based non-parametric probabilistic model for spatially interesting phenomena and connection links. The key opportunity lies in reducing the disjoint between application driven empirical and theoretical research that is currently commonplace in the building energy space.
\end{itemize}

% % \input{interconnect}

\section{Interconnect subsystems}
\label{sec:interconnect}
\noindent Buildings consist of multiple sub-systems such as networking, utility (electricity, water and gas) and security. Traditionally, many of these sub-systems are optimized for operation in isolation. However, since buildings are a unified ecosystem, optimal operations across the building would require understanding the relationships amongst these sub-systems and  interconnecting them. Previous work has shown that the sum of information obtained from each of these sub-system is greater than each one of them individually considered. We now discuss subsystem interconnection for occupancy based lighting and HVAC control.

\begin{figure}
  \centering
  \includegraphics[scale=0.19]{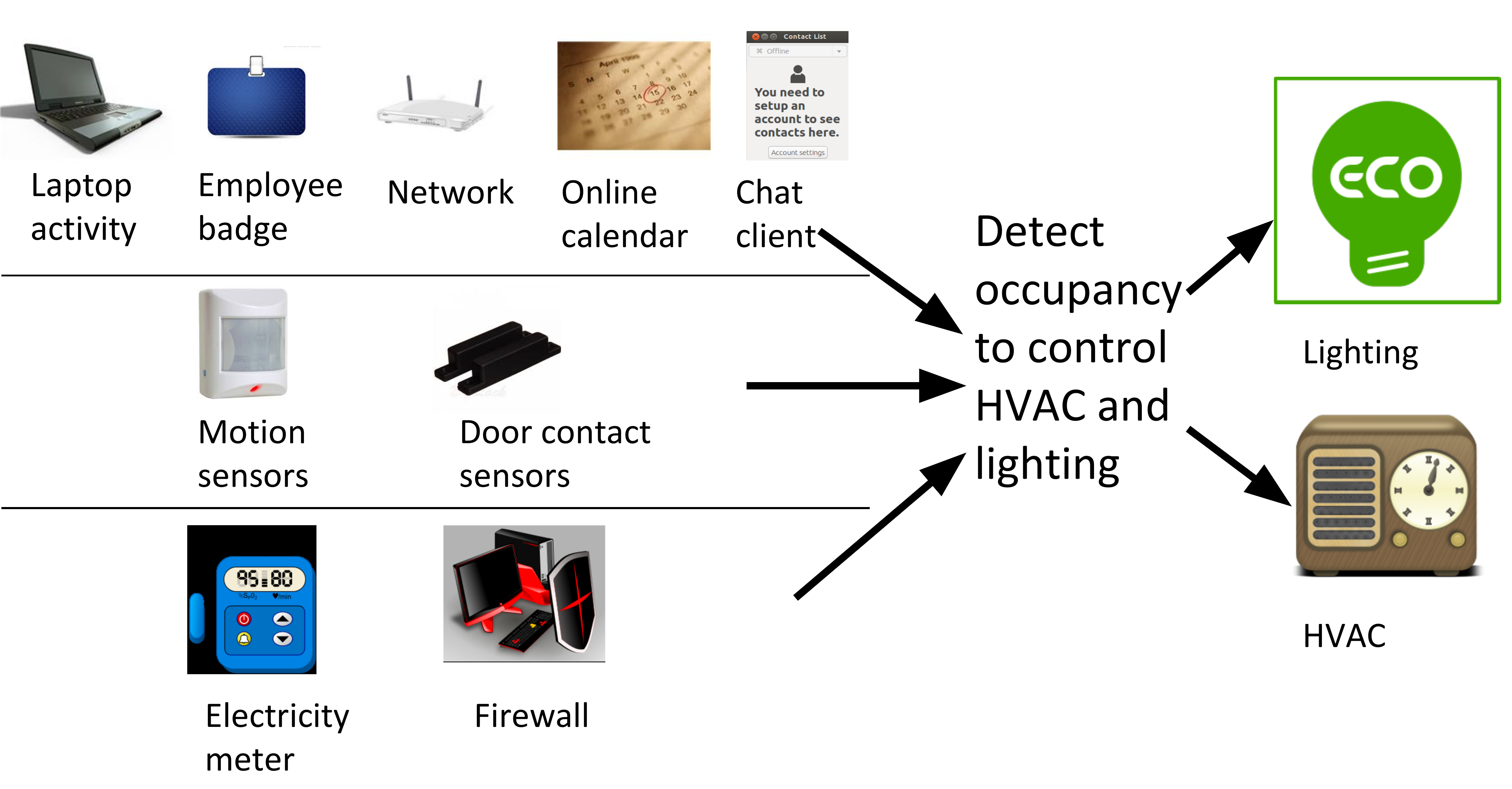}
  \caption{Different systems (top: using existing infrastructure and soft sensors~\cite{softgreen}, middle: using motion and door sensors~\cite{occupancy_ucsd} and bottom: using electricity meters and network firewall~\cite{granger}) interconnected to deduce occupancy for optimized lighting and HVAC control.}
  \label{fig:interconnect_sensing}
\end{figure}

As we shall discuss in \secref{sec:involve}, occupancy levels have a significant impact on building energy consumption. Previous studied have looked into occupancy inference using a variety of system interconnections. For instance, Thanayankizil et al.~\cite{softgreen} fuse information from networking equipment such as Wi-Fi access points with Office chat and meeting software for detecting occupancy for optimizing HVAC operations. On the same lines of using existing infrastructure, Kim et al.~\cite{granger} interconnect electricity data with firewall traffic to detect occupancy. Lu et al.~\cite{smart_thermostat} and Scott et al.~\cite{preheat} obtain occupancy patterns from Passive infrared sensors (PIR) and door sensors to control the home thermostat to conserve electricity in the absence of occupants. Similarly, Agarwal et al.~\cite{occupancy_ucsd} and Batra et al.~\cite{batra_issnip} use motion and door sensors for occupancy detection in office settings. On similar lines, Delaney et al.~\cite{lighting} discuss interconnecting motion sensor, light sensors and lighting control circuit for efficient lighting controls accounting for the availability of sunlight and occupant presence. \figref{fig:interconnect_sensing} shows various interconnected systems to detect occupancy for lighting and HVAC control.

Apart from subsystem interconnection for occupancy based control, there exists active interest in using the now ubiquitous smartphones for various energy applications, which we do not cover here in interest of space. Another active opportunity exists in exploiting the inter relationship between water and energy sub-systems in a building (discussed in \secref{sec:infer}).

Interconnection of various sub-systems has been made possible due to the fact that many systems can expose their data over IP, as discussed in \secref{sec:instrument}. Furthermore, various middleware systems~\cite{sensoract, building_depot, gsn} have been proposed in the recent past which greatly aid in collecting data from diverse sensors by writing wrappers and exposing the data in open formats. Many such middleware systems aim to expose buildings as a unified system of individual sub-systems, whereby, end user applications can be developed exploiting multiple sub-systems.

\subsection{Challenges and opportunities}
We now discuss some challenges and opportunities pertinent to interconnecting building sub-systems.
\begin{itemize}
\item \textbf{Complexity of writing applications:} Buildings are not currently programmable in a meaningful sense
because each building is constructed with vertically integrated, closed subsystems and without uniform abstractions to write applications against~\cite{boss,building_depot2}. Further, applications developed for one building are often not portable to other buildings due the vast differences in buildings and the underlying sensors and control systems. To mitigate this, Krioukov et al. propose BAS~\cite{bas}, a building application stack designed specifically to ease writing portable building applications which are agnostic of underlying hardware and the communication interfaces (such as Modbus, BACnet). They introduce software engineering principles to the building domain and allow applications to be built on top of a rich query interface. Haggerty et al. discuss Building operating system services (BOSS)~\cite{boss} which can leverage BAS query language and other features to develop portable fault-tolerant applications.
% design follows a layered approach where the lowermost layer sitting close to the hardware provides an interface to the hardware. On top of this sits the driver layer which provides interface for units such as fan which must share common features (eg. get and set speed). Rich queries are enabled by modeling the functional and spatial relationships. Finally applications are built on top of this stack. 
While the challenge of application portability has been addressed to some extent via BAS and BOSS, it remains an open opportunity to unify existing building energy applications to adhere to such stacked designs. Additionally, it must be highlighted that similar to general purpose software applications, this software-oriented view of buildings is prone to low quality and ad-hoc methodologies calling for development of building application quality metrics.

\item \textbf{Proprietary communication protocols:} 
%Application portability remains a challenge not only due to the differences in buildings and the involved sub-systems, but also due to the fact that many sensors and control end points use proprietary protocols for data transmissions and control. 
The task of writing hardware agnostic applications, as discussed above, is made complex due to several of these devices communicating using proprietary protocols. For each such end point, a custom wrapper has to be written on an end device which can accumulate the data and make it available in an open data format (or allow for control interface using common interface formats). Different building sub-systems may employ different vendor locked solutions making interconnection even more difficult. This calls for the development of open communication standards enabling decoupled networking stacks. Often this job is simplified by introducing a gateway device that converts the proprietary interface (such as Zwave) into IP interface. However, this adds up to the overall cost and maintenance. As previously discussed in \secref{sec:instrument}, the shift towards IP based communication can mitigate this problem.

\item \textbf{Inter department communication gap:} While different sub-systems may employ different vendor locked solutions for their custom application, a greater challenge in interconnecting sub-systems is that of the communication gap existing between different departments. The involvement of humans in this task makes it inherently difficult. While non technical in nature, this concern can limit optimal interconnections since individuals often have in-depth understanding about their own domain while interconnections would require understanding across different domains. As an example, HVAC systems typically are taught to mechanical engineering experts while the expertise of software stacks for data communication  and storage often remains with computer science experts. 

\item \textbf{Unstructured data:} Many of the sub-systems discussed above are managed by field experts. Significantly important data collected from these experts such as CAD layouts, notes from the respective departments need to be transformed into useful metadata. In the past, Krioukov et al.~\cite{bas} have used image processing techniques to extract chillers and other building sub-systems from a CAD drawing. In general, this exercise is both non-trivial and time consuming as the data being unstructured does not adhere to any fixed schema. To mitigate this requires establishment of schemas for buildings across these sub-systems, some of which have been studied in the past.\footnote{One such example is gbXML~\url{http://www.gbxml.org/aboutgbxml.php}} 

\item \textbf{Data and control access:} By interconnecting sub-systems, an application built on top can leverage data access and control across these sub-systems. However, each sub-system may need to comply to its specifications for data access and control. This challenge can be mitigated via establishment of rules for data and control access~\cite{sensoract,boss} for different applications and sub-systems. While theoretically feasible, it remains a complicated challenge to address, in context of the above stated challenges.

\end{itemize} 

% %\input{inferences}
\section{Inferred Decision Making}
\label{sec:infer}
\noindent Traditionally, data from within the system has been used for simple decision making and rule of the thumb approaches. However, once the systems are interconnected, there is a deluge of data coming from diverse systems. This data can be used to improve upon the decision making for optimal building operations. Inferred decision making can help in identifying inefficiencies, raising alerts and suggesting optimizations.  For example, earlier lighting control was performed in an automated way based on a fixed motion sensor timeout. However, this does not incorporate ambient light levels and the motion sensor timeout is static, as decided at the time of audit or installation. Based on interconnecting light sensor data with finer grainer occupancy data, previous systems~\cite{lightwise} have discussed possible energy savings by inferring optimal lighting levels. 
%Similarly, utility companies previously relied on customers' phone calls to detect power outages. They can now leverage smart meter data from different homes to quickly detect power outages and plan accordingly. Earlier, customers used to call the utility company seeking explanation for their bills. Now, due to the availability of fine grained temporal and spatial energy consumption data, inferences on top can help the customers to better understand the causes of inefficiency and electricity bills. The HVAC systems in buildings are often run on static schedules as decided by the facilities department. However, studying the relationship between temperature and electricity demand~\cite{Kavousian2013a, RicharddeDear2002}, building managers can now optimize HVAC operations.
In \tabref{tab:inferences} we summarize several similar optimizations made possible by inferred decision making across common building energy applications. We now discuss in detail an upcoming opportunity in building energy which is facilitated by the interconnection between energy an water.

\subsection{Water-Energy nexus} 
\noindent Water and energy are two critical indicators of sustainability in an organization~\cite{energy_water}. While we have largely discussed building energy, the inter-relationship between energy and water prompts a detailed discussion. The water-energy nexus\footnote{\url{http://en.wikipedia.org/wiki/Water-energy_nexus}} is defined as the relationship between energy and water encompassing the following two dimensions: i) energy usage in water treatment ii) water usage for energy generation. In the context of buildings, the water-energy nexus is an important topic encompassing inter-disciplinary research and has been attracting attention in the research community as evidenced by workshops such as the ACM Buildsys. We now discuss the water-energy nexus in two different settings: 1)large industrial campus; 2)residential apartment.

Kadengal et al.~\cite{energy_water} study the inter-relationships between energy and water inside a campus and answer the following questions based on the collected data: i) What are the trade-offs between the external water footprint of a campus and its internal energy footprint of water? ii) Are improvements in either footprint realizable in practice? iii) Does reducing the consumption of one water grade have more impact on the energy consumption than other water grades? and iv) Does rainwater harvesting help reduce a facility's energy footprint.

\begin{figure}
\includegraphics{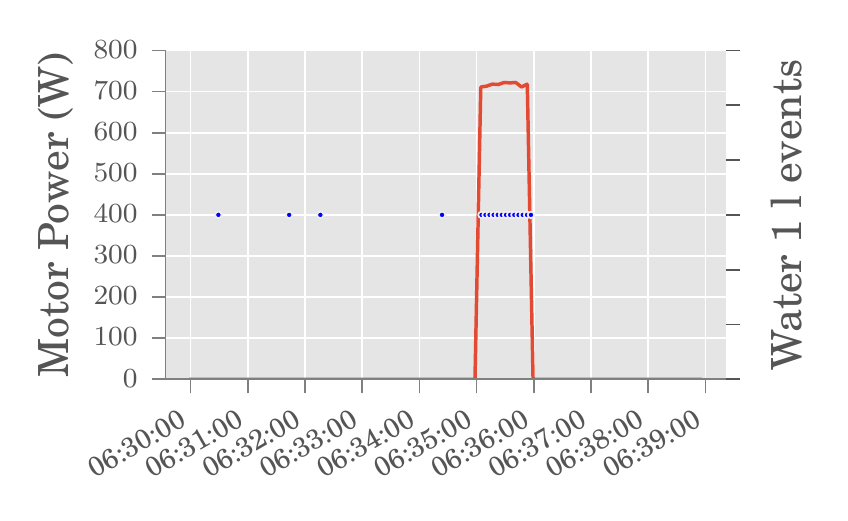}
\caption{Observing the energy water nexus in a residential apartment. The water motor increases the pumping rate at the cost of 1 horsepower}
\label{fig:motor}
\end{figure}

In residential settings, previous research~\cite{challenges_residential} has remarked on the fact that simultaneously observing both the electricity trace and the water trace can reveal finer information than when each of the system is analyzed in isolation. \figref{fig:motor} shows the water pump rate from a home in Delhi, India~\cite{iawe}. Due to low water pressure and non-availability 24X7, water motors are used to pump water to an overhead tank. This figure captures the motor's energy-water relationship. While turning on the water motor increases the pumping rate from 1 l/min to 20 l/min, it incurs an additional expense of 1 horsepower. In a different application, Prodhan et al. discuss HotWaterDJ~\cite{hot_water_dj} which is a system for smart water heating reducing the energy consumption of water heaters.

\begin{table}
\centering
\scriptsize
\begin{tabular}{p{1.3cm}p{2.5cm}p{3.5cm}}
\hline
\textbf{Applica-tion}&\textbf{Without inferred decisions}&\textbf{With inferred decisions}\\ \hline
Lighting control& Turn off lights after a threshold interval in which no motion has been detected& Fine control of light based on amount of ambient light and detailed occupancy information for light level control to match optimal conditions \\ \\
HVAC scheduling&Fixed 9 AM to 5 PM&Scheduling based on occupancy\\ \\
Utility power outage detection& Rely on phone calls from distressed consumers several minutes after an outage& Observe realtime data from smart meters and deploy corrective actions\\ \\
HVAC chiller temperature set point& Fix a setpoint as per one time audit& Correlate HVAC consumption with ambient temperature and working conditions, season and occupancy levels\\ \\
Utility consumer complaints for high pricing& Customers used to call the utility company and complain regarding extorbitant bills& Customers can view their consumption in realtime and set targets exceeding which they can be easily alerted\\ \hline
\end{tabular}
\caption{Common building energy applications which can be optimized by inferring decisions}
\label{tab:inferences}
\end{table}
%\noindent Data deluge in the building energy domain owing to collected data from instrumenting and connecting sub-systems opens up many new possibilities, which were deemed improbable in the past. 
%The scale of this problem makes the domain counted amongst the modern ``big data" problems. 
%In the recent past, availability of large quantities of data has enabled personalized recommendation based systems and unraveled newer possibilities in healthcare and astronomy domains. This rich amount of data can help make statistically sound inferences translating into control actions as opposed to following the conventional rule of thumb approach. It is expected that new advancements are similarly possible by inferences made on building energy data.

%We briefly present a few examples of inferences which can be made when such data is made available.

\subsection{Inference approach categorization}
\noindent Progressive research in the field of data science and building energy has led to newer inference approaches being developed to cater building energy applications. Comparison of these approaches is an application specific task. However, certain categorizations need to be studied for a fair comparison of these inference approaches. While this categorization is true across data science in general, we present a summary to understand the design space of inference algorithms in the context of building energy.
\begin{itemize}
\item \textbf{Centralized vs Distributed:} In centralized approaches the data resides and is processed on a single machine. The data deluge calls for a need to dwell into distributed systems where efficient inference and learning can be done via the use of distributed algorithms performed on data stored on different machines/nodes. We illustrate the difference between these two approaches via the problem of appliance fault detection. Traditionally, for solving this problem, sensor nodes measure various power features (e.g. active and reactive components) at the outlet level and relay them to a central controller. The central controller maintains a mapping between a node and the appliance it is connected to. The fault detection algorithm running at the central controller detects faults based on the data relayed by different nodes. However, since fault in an appliance is almost always independent of faults in other appliances, this problem has an inherent distributed characteristic. Ganu et al.~\cite{socket} take a distributed approach for this problem, whereby, each sensor node decides whether the connected appliance is faulty or not. This approach does not require data transmission to a central controller and can be also exploited for local control whereby the sensor node can individually indicate faulty operation.
\item \textbf{Supervised vs Unsupervised:} Supervised learning involves collection of ground truth/labeled data for training. As discussed in \secref{sec:instrument}, this can be expensive and non-trivial. Unsupervised approaches where learning does not involve labeled data and labels are discovered as opposed to explicitly being specified are more attractive. To illustrate the difference between the two categories, we consider the task of finding fixtures (both electricity and water) in a home. Typical supervised approaches involve instrumenting each fixture with a sensor for collecting labeled events (for example~\cite{watersense,viridiscope}). The training data is then used for learning classifiers. On the other hand, Srinivasan et al.~\cite{fixturefinder} use smart meter data along with data from already existing security systems for unsupervised event clustering and labeling for fixture discovery.
\item \textbf{Online vs Offline:} Another categorization of inference mechanisms which is widely prevalent in recommender systems and search engines is online versus offline models. Unlike offline models, online models can adapt to new data on the fly without the need to relearn the entire model from scratch. As the size of data increases, online models become more relevant. However, often online algorithms can not have a long term vision built using offline collected data and hence may need to take locally optimal decisions that may be globally non-optimal.
\end{itemize}
We believe that the optimal category of inference and learning approaches should be online, distributed and unsupervised algorithms. This is due to the challenges which exist in inferred decision making which we highlight now.

\subsection{Challenges and opportunities}

\begin{itemize}
\item \textbf{Collection of ground truth:} For quantifying improvements or effect of various proposed strategies, one needs to collect ground truth data. Furthermore, every new building on which evaluation is to be performed requires instrumentation. Due to this, scalability of the application is greatly hindered. The opportunity lies in developing easy to configurable and extensible building simulators (or interfaces to existing ones. Recently, Bernal et al. presented MLE+~\cite{mleplus}) which leverages the high-fidelity building simulation capabilities of EnergyPlus and the scientific computation and design capabilities of Matlab for controller design. The presence of such simulators can also mitigate the challenge previously faced in the research community when two different strategies were performed under different settings and could not thus be directly compared.
\item \textbf{Need for online and tractable algorithms:} Computationally intractable algorithms for performing inference and learning scale poorly with the size of the data. As discussed in \secref{sec:building_energy_deluge}, data size in building energy domain is increasing rapidly. Thus, the traditional offline intractable algorithms need to be complemented with their tractable and online equivalents for timely actionable insights.  
\item \textbf{Testing in real environments:} Even though effective simulators may be developed, testing the validity of the inference produced in the real world is hard to realize. This is due to the complications arising out of closing the loop requiring interface with the existing control system.

\end{itemize}

%Having discussed a few challenges and opportunities, we now discuss in detail a big opportunity in this field, namely the Water-Energy nexus.

% % \input{involve}
\section{Involve occupants}
\label{sec:involve}
\noindent 
%Inferred decisions can translate into savings via control action in two ways. Firstly, by involving occupants and secondly by intelligent operations. In this section, we discuss about involving occupants.

\noindent Modern buildings are designed to provide occupant comfort, security and energy efficiency, among others. Owing to these design goals, buildings form a complex ecosystem consisting of physical, cyber and human components and their interactions~\cite{satyanarayanan}. Occupants form an integral part of these energy efficient buildings ecosystem and their presence makes this ecosystem even more complex as occupants have individual preferences. Hsu et al.~\cite{hbci} coin the term Human-Building-Computer-Interaction (HBCI)] to describe the inter-relationships between human beings, buildings and computers. \figref{fig:hbci} shows the HBCI as proposed by the authors. 

Traditional systems missed out on one or more component in the HBCI framework while trying to aim for optimal environment. The occupants and their preferences were mostly not taken into account and the decisions for energy efficiency were taken centrally. However, the HBCI framework helps in formalizing the optimization problem involving human comfort and energy savings. Occupants' behavior plays a significant factor in this optimization problem. Previous literature~\cite{energy_buildings_survey} suggests that energy unaware behavior can add upto one-third to a building's energy performance. On the other hand, Darby et al.~\cite{darby_2006} suggest that providing the occupants feedback on energy usage can provide up to 15\% savings. Thus, both occupant comfort and energy efficiency can be considered as joint goals of a smart building and an optimal sweet point balancing their trade offs is desired.

\subsection{HBCI}
\begin{figure}
\centering
\includegraphics[scale=0.3]{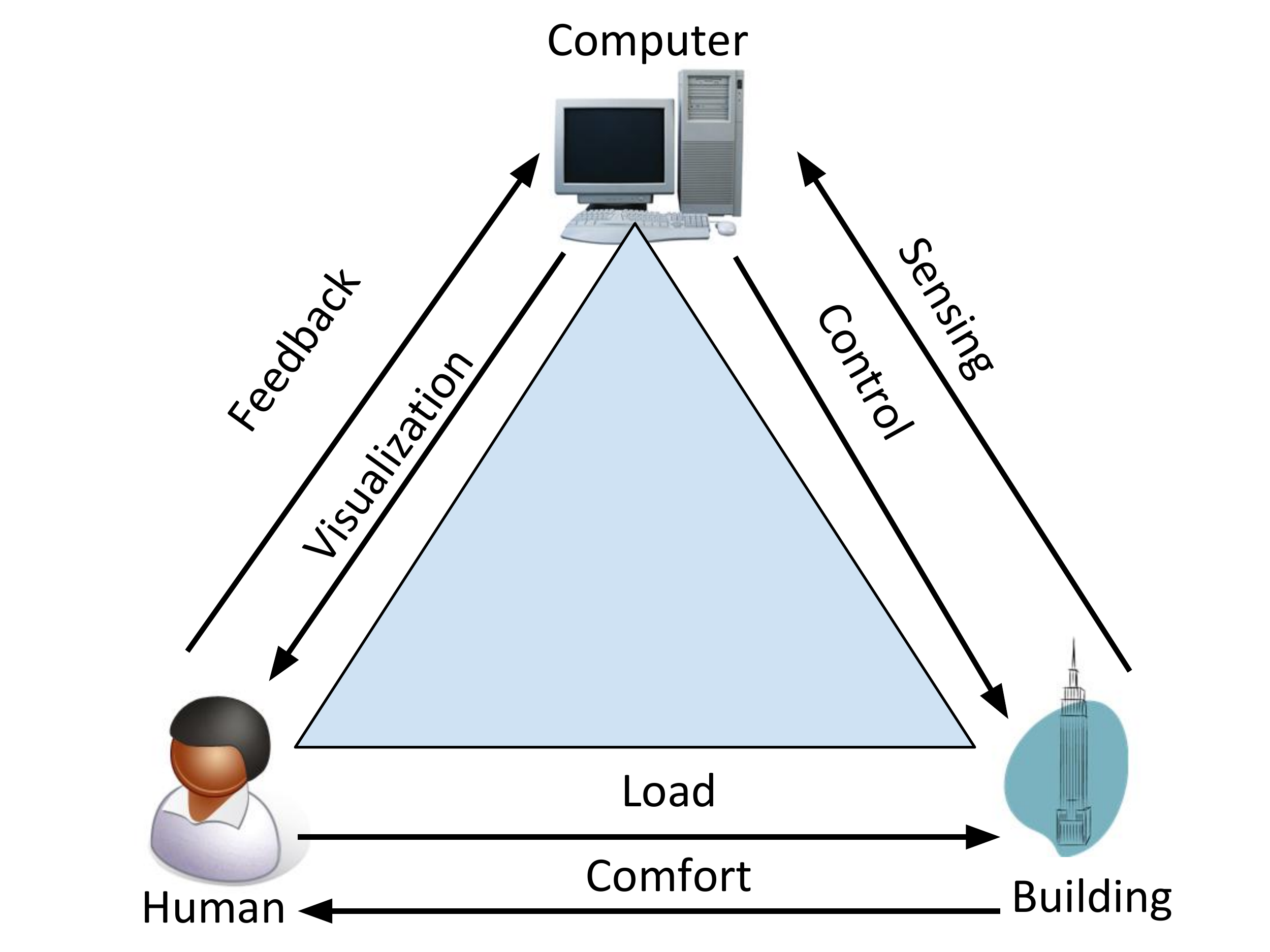}
\caption{Human-Building-Computer interaction (HBCI)-architecture to bridge the gap between the digital and physical worlds [Figure adapted from~\cite{hbci}]}
\label{fig:hbci}
\end{figure}
We now discuss in detail HBCI and how the different interactions relate with energy efficiency and the other Is. As per \figref{fig:hbci}, there are four primary occupant interactions in the HBCI system which are as follows:
\begin{enumerate}
\item \textbf{Occupants provide feedback for better computation:} \label{sec:feedback_building}Building occupants can provide valuable information for improving efficiency (feedback in \figref{fig:hbci}) in a variety of ways. Previous studies have instrumented human bodies for a variety of applications such as healthcare~\cite{body_1,body_2}. However, human wearable sensors and the collected data thereof, are highly intrusive and are not included in the scope of this paper. Building occupants can also indirectly provide their location in the building. Several studies in the past have used cell phone data to localize an occupant in the building~\cite{radar,indoor_1,indoor_2}. This can greatly help in activity recognition, culminating in accurate energy apportionment~\cite{apportionment} and in several other energy saving measures such as modifying HVAC parameters to account for occupancy and turning off lights where no one is present~\cite{reduce}. This interaction between occupants and the computing world rs can be seen as an extension of \textbf{optimal instrumentation}, where the occupants and their possessions (such as cell phones) can themselves act as sensors reducing the necessity for dense sensing deployments.
\item \textbf{Computation to provide feedback and suggestions to occupants:} Personalized recommendation systems have been well studied in the entertainment sector. A similar opportunity lies in the building energy domain as well. Energy dashboard~\cite{energy_dashboard} systems have been proposed in the past to make the users aware of their energy footprint. Providing personalized energy consumption information has also been made possible via energy apportionment strategies~\cite{apportionment,granger}. It is believed that providing occupants with detailed feedback on their consumption can help reduce their energy footprint by induced behavioral changes~\cite{bidgely_2013,darby_2006}. Recently, Batra et al.~\cite{iawe} and Ganu et al.~\cite{socket} discuss systems for appliance fault detection. When appliance faults are detected, it is upto the occupants to take proper action when informed about possible optimizations. 

%Occupants can act on such detailed information and recommendation to save energy.

\item \textbf{Occupants load buildings:} Occupants occupy the physical space in the building and utilize various resources. Several occupant actions usually translate into energy consumption inside buildings. For instance, in a residential building, occupants may use different electrical appliances such as television or oven. In commercial buildings, occupants may interact with servers and other resources. The occupancy on the building often directly relates with the energy consumption of the building. Decisions may be inferred to decide the optimal operating conditions in a building based on occupancy load. Such schemes ensure that occupant comfort is not compromised and at the same time energy efficient measures are considered.

\item \textbf{Expecting comfort from building:} As discussed earlier in this section, control actions were previously taken mostly at a central facility level and were totally agnostic of individual preferences. However, by providing feedback to the building management system about their preference, individuals can better control their environment translating to both higher comfort and increased energy efficiency. Previous research (personalized lighting~\cite{lighting} and participatory HVAC control~\cite{thermovote}) has proposed systems for participatory and individual HVAC and lighting control.
\end{enumerate}

\subsection{Challenges and opportunities}
We now discuss some challenges and opportunities in involving occupants in the context of building energy.
\begin{itemize}
\item \textbf{Privacy concerns:} Empowering the occupants with more information about their activities of daily life and consequently impact on energy consumption has significant privacy concerns. This may greatly reduce the involvement of occupants who may feel threatened about the amount of data being collected and the potential information that can be inferred. This reaction may be similar to the reaction of users to internet companies collecting a lot of user data for targeted advertisement. Several activities can be easily inferred even from a whole home smart meter. Further, when seen in conjunction with other sensors commonly present in buildings, much finer grained information can also be revealed. Molina et al.~\cite{private_memoirs} address this challenge via a privacy enhancing scheme for preserving metering goals of a smart meter while protecting occupant information. Thus, there is a need to develop privacy preserving architectures to allow meeting the goal of building energy monitoring while still maintaining occupant privacy~\cite{sensoract}. 
\item \textbf{Indifferent occupant attitude:} Previous research ~\cite{granger} suggests that if the occupants do not directly pay for their electricity bills, such as in commercial offices and academic institutes, they may not be enthusiastic towards electricity savings. Under such circumstances, indifferent occupant attitude remains a big challenge and needs to be accounted for in future work.
It is also possible that occupants may initially show interest and may provide regular feedback for improving the system. However, with time the users may fall back to their usual schedules and thus, sustained occupant engagement remains an open challenge. Novel human computer interaction (HCI)~\cite{novel} techniques can be used for better occupant engagement.
\end{itemize}

% % \input{intelligent}

\section{Intelligent Operations}
\label{sec:intelligent}
\noindent All of optimal sensing from interconnected subsystems, resulting in decisions inferred from the rich dataset while involving the occupants will result in intelligent operations. Such intelligent operations already exist but miss out one or more aspects discussed previously i.e. either they do not involve the occupants or are taken at a sub-system level without accounting for data from other sub-systems in operation. Realizing energy efficiency requires striking the perfect match between the five Is. For instance, a system having advanced control capabilities but poor inference capabilities is unlikely to result in energy efficiency. Likewise, a system having advanced inference capabilities, but poor control capabilities will only be able to propose efficiency and not realize it.

We base our discussion in this section by discussing a simplified view of a smart grid (which excludes the role of government and regulators) and the role of buildings in it. Our simplified smart grid setup has three key activities: 1) generation 2) transmission and 3) consumption. Of these three activities, buildings were traditionally involved only in electricity consumption. However, buildings are increasingly equipped with generation capabilities through renewable resources (e.g. solar). Since the renewables involve generation that is typically weather dependent and hence is time varying, such sources bring forth an interesting optimization perspective - if the production is more than what can be consumed, how best to utilize the remaining excess generated power. These topics are usually studied under the umbrella field of automated demand response which we briefly discuss now.

Demand response is defined as - ``Changes in electric usage by end-use customers from their normal consumption patterns in response to changes in the price of electricity over time, or to incentive payments designed to induce lower electricity use at times of high wholesale market prices or when system reliability is jeopardized"~\cite{adr}. While generally energy efficiency is considered to be closely related with energy conservation, studying demand response allows us to present an alternative understanding. We had discussed earlier in \secref{sec:building_energy_deluge} that commercial buildings are billed on time-of-day based pricing. This is due to the fact that over and above the average load, the utilities have to expend a lot of resources and money to cater to peak demand. Thus, in order to encourage consistent load profiles (with minimum time variation), the utilities introduce variable pricing- non peak hours have cheaper electricity rates as compared to peak hours. Building energy research community and the industry has looked into this problem and has come up with several ways of peak demand flattening. We present two such approaches as follows.
%\noindent While energy data and inferences may help identify core areas to improve energy efficiency, they will bear fruit only when acted upon. We believe that these control actions can come in two forms:
%i) Control via human in the loop and ii) Automated control

%Controlling via human in the loop was discussed previously in \secref{sec:involve}. In this section, we discuss automated control operations also called intelligent operations for energy efficiency in buildings. 

Firstly, we discuss SmartCap~\cite{smartcap}, a scheme based on the well known scheduling algorithms used in real time operating systems. The authors discuss dividing electrical loads into two categories- background loads which can run without human intervention (such as refrigerator, air conditioner) and interactive loads (such as television, stove, microwave, lighting). Since scheduling interactive loads may interfere with daily life of the occupants, only background loads are rescheduled to alleviate peak demand. Each load is modeled with a simple slack, or time required for it to complete its operations (e.g. for a refrigerator the operation goal may be to reach the set temperature within the next hour). 

In contrast, companies such as Stem have proposed battery storage schemes for reducing peak consumption. The battery is charged from the grid during the off-peak period and is discharged to lower the demand from the grid during peak period, thereby, reducing the overall peak demand and flattening the power draw. The company claims to reduce up to 20\% of the overall bills by this storage based mechanism to lower peak demand. 
%Similar practices have been used in the state of California where water is pumped using off-peak electricity and the potential energy stored in the pumped up water is used to cater to the additional demand during peak hours. 
Integration of local generation via renewables (e.g. solar) with such batteries remains an interesting open possibility, which becomes more difficult due to the time varying nature of electricity produce.

While theoretically such operations seem feasible, they require an elaborate control system setup. In the case of SmartCap like approaches, a control system needs to interact with the inference engine which decides the slack for each appliance. Correspondingly, the operating condition for each appliance has to be decided. Subsequently, the control also has to abide by certain limits such as limited the rate of appliance state toggling to prevent damage. In the second case (Stem), where additional batteries are provisioned, the control system must have the capability of efficiently changing the power supply from the grid to the battery and vice versa. Due to the existence of such complications and risks, research systems in the past have usually only presented proof of concept intelligent operations as opposed to realizing these in the real world.

%
%
%These two different ways of alleviating peak demand present wherein all of the other four Is have complemented to culminate in intelligent operations involving control actions.

\subsection{Challenges and opportunities}
We now discuss some of the challenges and opportunities in performing intelligent operations.
\begin{itemize}
\item \textbf{Significant up front costs:} To realize the full potential of such intelligent operations, there may be significant up front cost. In the case of battery based solutions, one needs to invest heavily in managing additional equipment. In case of approaches such as SmartCap one needs to add control capability to existing monitoring equipment. Additionally, each appliance must be individually instrumented. 

\item \textbf{Security breach:} To facilitate schemes such as SmartCap, appliance owners can take a step towards making their appliances more IoT ready. Appliances can allow their configuration over standard HTTP based interfaces. However, taking a step towards controlling appliances for peak demand and especially more so over standard interfaces like HTTP, exposes them to cyber attacks. Since most of the end devices usually have low computing and memory capabilities, they are limited in terms of computational capabilities needed for advanced encryption systems. Thus, allowing intelligent operations while maintaining security aspects in resource constrained end device remains a challenge to be addressed in the future.

\item \textbf{Complex control environment:} Controlling operations of a building encompasses a variety of systems needing highly sophisticated controls. Thus, introducing optimizations into such a complex system is highly risky. While simulations may be effectively used, the real world may bring up unforeseen challenges. Moreover, the controls which may work well in one building, may miserably fail in another. An analogous example may be the crashing of Ariane V spacecraft~\footnote{\url{http://www.around.com/ariane.html}}. The spacecraft used the software from its previous version, but had upgraded hardware. The faster velocity attained by the new hardware caused numerical overflow and the control system of the spacecraft failed causing auto destruction. While the empirical research may promise exciting results, it could miss one or more cases which could cause a disaster similar to that of Ariane V. This calls for the development of theoretical proofs of valid and risk averse control and operation and their integration into existing simulators. 
\end{itemize}

% % \input{applications}

\section{Applications}
\label{sec:applications}
\noindent Having discussed each of the 5 Is in detail, we now present a summary study of recent applications in the building energy domain in this section. Most of these applications span across the 5 Is as discussed above and are summarized in Table \ref{table:applications}.

\begin{table*}[!t]
\centering 
\scriptsize
\begin{tabular}{p{2cm}p{0.5cm}p{2cm}p{2.1cm}p{2.1cm}p{2.1cm}p{2.1cm}p{2.1cm}}
\hline
\textbf{Authors} & \textbf{Year} & \textbf{Key \newline application} & \textbf{Instrumentation} & \textbf{Interconnect} & \textbf{Infer} & \textbf{Involve \newline occupants} & \textbf{Intelligent \newline operations}\\ \hline

Delaney et al. ~\cite{lightwise} &  2009 & Lighting control & Light sensor, motion sensor & Light and motion sensor & Infer savings when unoccupied or light above threshold &- & Suggests integrating energy saving algorithm into control system\\  \hline 

Hay et al. ~\cite{apportionment} & 2009 & Energy apportionment& Electric meters, diary entries, door security, appliance clamp meters& Combine electric meters, door security and diary entries  & Apportion individual energy usage by understanding context and apportionment policy & Provide detailed apportioned energy&-\\ \hline 

%Agarwal et al.~\cite{energy_dashboard} & 2009 &\\

Lu et al.~\cite{smart_thermostat} & 2010 & Smart thermostat &Thermostat, motion and door &Determine occupancy from combining door and motion data & Optimized temperature setting & Occupants establish ground truth by taking notes& Pro-actively control thermostat set point\\ \hline

Kim et al. ~\cite{granger}  & 2010 & Energy apportionment & Circuit panel electricity, firewall & electricity and firewall & Find the causality between IP and circuit level energy consumption & Provide apportioned energy usage and user feedback & -\\ \hline

Weng et al.~\cite{plug_load}& 2011& Demand response& Smart plugs, door, motion sensors& door and motion & Load shedding potential and occupancy & Enlist occupant priority for load actuation & Automatically turn off loads in response to demand\\ \hline

Srinivasan et al. ~\cite{watersense} & 2011 & Water disaggregation& Household flow meter, motion sensors& Combine motion, room metadata and flow& Bayes net to label flow events& Provide room and fixture information; may act upon data from feedback& Provide feedback to occupants\\ \hline
 
Krioukov et al. ~\cite{lighting}&2011 & Personalized automated lighting& access existing BACNet&Lighting control and monitoring and electricity over BACNet&Estimated energy saving&Occupants have access to web application to control and monitor light usage&Control light based on human input and some user thresholds\\ \hline

Schoofs et al. ~\cite{copolan} & 2011 & Non-invasive occupancy profiling& Electricity meters, VLAN& VLAN, electricity& Correlate VLAN and electricity consumption&Collect occupant entry and exit times&Dynamically optimize HVAC based on occupancy inferred from VLAN\\ \hline

Kamthe et al. ~\cite{energy_auditing}& 2011 & Energy auditing & camera sensors&-&Tune previous model from similar layout building according to occupancy information measured for small duration&-&Optimize baseline building strategies\\ \hline

Prodhan et al.~\cite{hot_water_dj} & 2012 & Smart water heating & Temperature and water flow sensors & Water flow and temperature to find & Learn temperature and lag time for fixtures& Occupants configure comfort knobs for different fixtures & Choose fixture temperature and delay\\ \hline

Thanayankizil et al. ~\cite{softgreen} & 2012 & Opportunistic context sources & Soft sensors-WiFi, access cards & Fuse different context sources& Employee location and room occupancy & Access card swipe, laptop monitoring& HVAC optimization \\ \hline

Barker et al. ~\cite{smartcap} & 2012 & Peak load reduction& Smart meter, Insteon & - & Slack time for appliances & - & Schedule background loads while maintaining operation requirements\\ \hline

Rogers et al. ~\cite{myjuolo} & 2013 & Personalized heating advice & Temperature logger & Indoor temperature and outside temperature from weather station& Thermostat model and savings from changing operations& Provide actionable feedback to occupants \\ \hline

Chen et al.~\cite{niom_umass} and Kleiminger et al.~\cite{niom_eth} & 2013 & Occupancy prediction from smart meter data & Smart meters & - & Infer occupancy from various power features obtained from smart meters & - & Suggested: integrate with energy efficiency solutions relying on occupancy\\ \hline

Beltran et al.~\cite{thermosense} & 2013 &Occupancy sensing&thermal and motion sensors&Motion sensor triggers thermal sensors when motion detected&Use classifiers and filtering for predicting occupancy&-&Optimize HVAC control based on occupancy\\ \hline

Saha et al. ~\cite{energylens} & 2014&Energy Apportionment&Smart Meters and smartphone (WiFi, audio)&Combines smart meter data with smartphone sensor data&What,whom, where, when of appliance usage&Provide detailed apportioned energy information&Suggests using context information from smartphones with smart meter data\\

\hline

\hline
\end{tabular}
\caption{Illustration of multiple building energy applications and how they span across the 5 Is}
\label{table:applications}
\end{table*}

We specifically discuss in detail about the well studied problem of non-intrusive load monitoring (NILM) and discuss how it spans across the 5 Is.

\begin{figure*}
\centering
\includegraphics[scale=0.9]{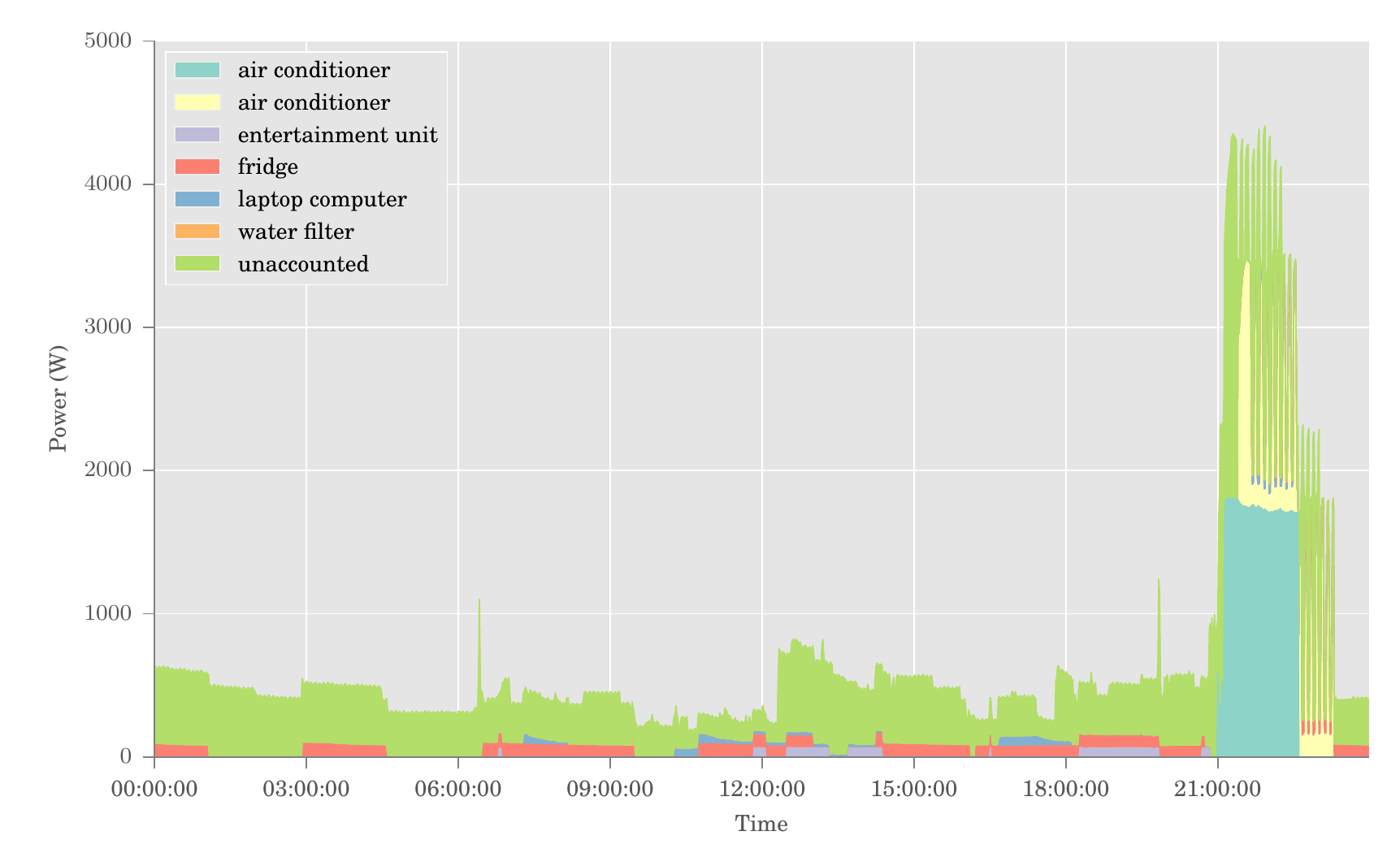}
\caption{Disaggregated consumption [Best viewed in color]}
\label{fig:nilm}
\end{figure*}

\subsection{Non-intrusive load monitoring}
\label{sec:nilm}
\noindent Non-intrusive load monitoring (NILM) or energy disaggregation is the process of breaking down the energy measurement observed at a single point of sensing into constituent loads~\cite{hart_1992}. \figref{fig:nilm} shows disaggregated consumption across a day for different appliances from the iAWE~\cite{iawe} data set. In 2006, Darby et al.~\cite{darby_2006} suggested that providing detailed electricity information feedback to end users can lead to 5-15\% electricity usage reduction by behavioral change. Recently, Chakravarty et al.~\cite{bidgely_2013} performed a study across more than 300 users in California and observed mean reduction of approx. 15\% when disaggregated information and real time electricity information is provided to end users.

With the advent of smart metering infrastructure, as discussed in \secref{sec:introduction}, NILM has had a lot of renewed interest. A number of startups\footnote{Such as Bidgely, Neurio, Plotwatt} which provide itemized electrical usage to households over cloud based services have emerged recently. The recent interest has also led to the release of data sets meant to aid NILM research. Of these REDD~\cite{redd}, BLUED~\cite{blued}, AMPds~\cite{ampds}, UK-DALE~\cite{ukpd}, ECO~\cite{eco}, Sust-data~\cite{sust}, GREEND~\cite{greend}, COMBED~\cite{combed}, BERDS~\cite{berds} and Pecan~\cite{pecan} have been specifically released for NILM like applications. Apart from these data sets, other data sets such as Tracebase~\cite{tracebase}, HES~\cite{hes}, SMART*~\cite{smart} and iAWE~\cite{iawe} can be used for several other building applications, including NILM. We now discuss the NILM problem across the five Is. 

\noindent \textbf{Instrument optimally: }NILM involves breaking down the aggregate meter data measured at a single point into constituent appliances or sub-meters lower in the electrical tree. Thus, instrumentation for NILM involves measuring electrical parameters at both the aggregate and the sub-metered level (desired for ground truth). The aggregate power readings are typically measured using smart meters. In special cases, when high frequency data (more than 1 KHz) is required, sophisticated data acquisition systems (DAQs) are used. Sub-metered power data is typically measured either at the circuit level using current transformer (CT) based sensors or at appliance level using appliance sensors. 

\noindent \textbf{Interconnect sub-systems:} While in the classic NILM problem, interconnection across different data streams is not studied, some studies use the extra information from other modalities to improve disaggregation. Previous work~\cite{RicharddeDear2002, Kavousian2013a} indicate the correlation between energy usage and external temperature. Berges et al.~\cite{multimodal} correlate occupancy sensors with electrical data to identify potential savings in unoccupied rooms. Similarly, other ambient sensors can also be interconnected with electrical sensors to obtain additional insights into the disaggregation problem~\cite{fixturefinder}.

\noindent \textbf{Inferred decision making:} The vanilla use case of the NILM implementation provides an itemized breakdown of the electrical load. This inference problem can be viewed as an inverse classification problem. Previous work has related this problem to source separation which is a well studied problem in sound processing. Hart et al.~\cite{hart_1992}, in their seminal work on NILM, proposed a simplistic combinatorial based and a simple edge detection based NILM algorithm. Both these algorithms form the foundation behind many of the state-of-the-art algorithms. Markovian analogues of combinatorial optimization formulation led to factorial hidden Markov~\cite{fhmm} among other hidden Markov models. Several NILM approaches have been proposed in the recent past~\cite{parson_2012, johnson_2013, rahayu_2012, batra_2013, kolter_2012, anderson_2012} and a rich overview has been captured in several recent work as well~\cite{survey_nilm, holy_grail, zeifman_2011,nilmtk}. It must be pointed that most of the prior research includes \textbf{supervised} methods in a \textbf{centralized} \textbf{offline} setup.

\noindent \textbf{Involve occupants:} As discussed earlier in this section, providing itemized feedback to end users has shown to reduce their end consumption. Thus, the occupants can be involved by not only providing them with itemized billing, but also providing actionable suggestions on top.
Another interesting occupant involvement may arise from devising novel techniques for ground truth collection, which eliminates the need for appliance metering.

\noindent \textbf{Intelligent operations:} Real time control actions to save electricity based on disaggregated information require complicated interactions with the control systems. Current literature is thin on the aspect of automated control beyond the usual feedback that NILM systems provide.

\subsubsection{Challenges and opportunities}
In this section we briefly mention some of the challenges and opportunities in NILM research.
\begin{itemize}
\item \textbf{Need for extensive deployment:} For certifying an algorithm on a previously unseen home, appliance level data must be collected from that home. 
\item \textbf{Computationally expensive approaches:} Many approaches are computationally expensive and thus model only the high energy consuming appliances. Owing to this, detailed information about low energy consuming appliances is often not made available. The computationally intractable approaches also limit the application in the real world setting.
\item \textbf{Supervised methods require sub-metered data:} Supervised NILM approaches train on the sub-metered data and create a model for each appliance. This requires ground truth instrumentation, motivating the need for development of novel unsupervised learning mechanisms.
\end{itemize}

% %\input{conclusions}

\section{Conclusions and Future work}
\label{sec:conclusions}
\noindent Optimizing building energy usage remains an area of concern in light of dwindling natural resources. Due to this concern, efforts are being concentrated to develop an understanding into buildings. These efforts have led to a deluge of data coming from a variety of sensors. This availability of data is changing the way we consume our electricity information. In this paper we highlighted some of the applications enabled by this data availability, such as early power outage detection, peak load reduction, electricity consumption reduction. 

Based on our literature survey of research in building energy domain, we identified five crust areas enabling data centric energy efficient buildings:
i) instrument optimally ii) interconnect sub-systems iii) inferred decisions iv) involve occupants and v)intelligent operations. Across each of these five areas we present the state-of-the-art, the core challenges and the opportunities in the field. Finally, we categorize different building energy applications as per these five Is and discuss non-intrusive load monitoring, a well studied problem in building energy domain, in greater detail.

%We discuss various applications spanning across each of these five crust areas. These range from developing smart appliances to developing mechanisms for demand response. We present a detailed analysis of NILM, which is a specific application cutting across these five areas. Finally, we present some audacious goals enabled by pervasive technologies spanning across these crust areas.

\bibliographystyle{abbrv}
{\scriptsize
\vspace{3pt}
\bibliography{reference}

\begin{thebibliography}{100}

\bibitem{gsn}
K.~Aberer, M.~Hauswirth, and A.~Salehi.
\newblock The global sensor networks middleware for efficient and flexible
  deployment and interconnection of sensor networks.
\newblock {\em Ecole Polytechnique Fdrale de Lausanne (EPFL), Tech. Rep.
  LSIR-REPORT-2006-006}, 2006.

\bibitem{occupancy_ucsd}
Y.~Agarwal, B.~Balaji, R.~Gupta, J.~Lyles, M.~Wei, and T.~Weng.
\newblock Occupancy-driven energy management for smart building automation.
\newblock In {\em Proceedings of the 2nd ACM Workshop on Embedded Sensing
  Systems for Energy-Efficiency in Building}, pages 1--6. ACM, 2010.

\bibitem{building_depot}
Y.~Agarwal, R.~Gupta, D.~Komaki, and T.~Weng.
\newblock Buildingdepot: an extensible and distributed architecture for
  building data storage, access and sharing.
\newblock In {\em Proceedings of the Fourth ACM Workshop on Embedded Sensing
  Systems for Energy-Efficiency in Buildings}, pages 64--71. ACM, 2012.

\bibitem{energy_dashboard}
Y.~Agarwal, T.~Weng, and R.~K. Gupta.
\newblock The energy dashboard: improving the visibility of energy consumption
  at a campus-wide scale.
\newblock In {\em Proceedings of the First ACM Workshop on Embedded Sensing
  Systems for Energy-Efficiency in Buildings}, pages 55--60. ACM, 2009.

\bibitem{sn_survey_1}
I.~F. Akyildiz, W.~Su, Y.~Sankarasubramaniam, and E.~Cayirci.
\newblock A survey on sensor networks.
\newblock {\em Communications magazine, IEEE}, 40(8):102--114, 2002.

\bibitem{anderson_2012}
K.~Anderson, M.~Berges, A.~Ocneanu, D.~Benitez, and J.~Moura.
\newblock Event detection for non intrusive load monitoring.
\newblock In {\em Proceedings of 38th Annual Conference on IEEE Industrial
  Electronics Society}, pages 3312--3317, 2012.

\bibitem{blued}
K.~Anderson, A.~Ocneanu, D.~Benitez, D.~Carlson, A.~Rowe, and M.~Berg{\'e}s.
\newblock Blued: A fully labeled public dataset for event-based non-intrusive
  load monitoring research.
\newblock In {\em Proceedings of 2nd KDD Workshop on Data Mining Applications
  in Sustainability}, pages 12--16, Beijing, China, 2012.

\bibitem{sensoract}
P.~Arjunan, N.~Batra, H.~Choi, A.~Singh, P.~Singh, and M.~B. Srivastava.
\newblock Sensoract: a privacy and security aware federated middleware for
  building management.
\newblock In {\em Proceedings of the Fourth ACM Workshop on Embedded Sensing
  Systems for Energy-Efficiency in Buildings}, pages 80--87. ACM, 2012.

\bibitem{autogrid_data_deluge}
Autogrid.
\newblock The energy data deluge, 2013.

\bibitem{radar}
P.~Bahl and V.~N. Padmanabhan.
\newblock Radar: An in-building rf-based user location and tracking system.
\newblock In {\em INFOCOM 2000. Nineteenth Annual Joint Conference of the IEEE
  Computer and Communications Societies. Proceedings. IEEE}, volume~2, pages
  775--784. Ieee, 2000.

\bibitem{adr}
V.~M. Balijepalli, V.~Pradhan, S.~Khaparde, and R.~Shereef.
\newblock Review of demand response under smart grid paradigm.
\newblock In {\em Innovative Smart Grid Technologies-India (ISGT India), 2011
  IEEE PES}, pages 236--243. IEEE, 2011.

\bibitem{smart}
S.~Barker, A.~Mishra, D.~Irwin, E.~Cecchet, P.~Shenoy, and J.~Albrecht.
\newblock Smart*: An open data set and tools for enabling research in
  sustainable homes.
\newblock In {\em Proceedings of 2nd KDD Workshop on Data Mining Applications
  in Sustainability}, Beijing, China, 2012.

\bibitem{smartcap}
S.~Barker, A.~Mishra, D.~Irwin, P.~Shenoy, and J.~Albrecht.
\newblock Smartcap: Flattening peak electricity demand in smart homes.
\newblock In {\em Pervasive Computing and Communications (PerCom), 2012 IEEE
  International Conference on}, pages 67--75. IEEE, 2012.

\bibitem{hitchhikers_wsn}
G.~Barrenetxea, F.~Ingelrest, G.~Schaefer, and M.~Vetterli.
\newblock The hitchhiker's guide to successful wireless sensor network
  deployments.
\newblock In {\em Proceedings of the 6th ACM conference on Embedded network
  sensor systems}, pages 43--56. ACM, 2008.

\bibitem{batra_issnip}
N.~Batra, P.~Arjunan, A.~Singh, and P.~Singh.
\newblock Experiences with occupancy based building management systems.
\newblock In {\em Intelligent Sensors, Sensor Networks and Information
  Processing, 2013 IEEE Eighth International Conference on}, pages 153--158.
  IEEE, 2013.

\bibitem{batra_2013}
N.~Batra, H.~Dutta, and A.~Singh.
\newblock {INDiC: Improved Non-Intrusive load monitoring using load Division
  and Calibration}.
\newblock In {\em International Conference of Machine Learning and
  Applications}, Miami, FL, USA, 2013.

\bibitem{iawe}
N.~Batra, M.~Gulati, A.~Singh, and M.~B. Srivastava.
\newblock {It's Different: Insights into home energy consumption in India}.
\newblock In {\em Proceedings of the Fifth ACM Workshop on Embedded Sensing
  Systems for Energy-Efficiency in Buildings}, 2013.

\bibitem{nilmtk}
N.~Batra, J.~Kelly, O.~Parson, H.~Dutta, W.~Knottenbelt, A.~Rogers, A.~Singh,
  and M.~Srivastava.
\newblock {NILMTK: An Open Source Toolkit for Non-intrusive Load Monitoring}.
\newblock In {\em ACM e-Energy}, Cambridge, UK, 2014 (in press).

\bibitem{combed}
N.~Batra, O.~Parson, M.~Berges, A.~Singh, and A.~Rogers.
\newblock A comparison of non-intrusive load monitoring methods for commercial
  and residential buildings.
\newblock {\em arXiv:1404.3878}, 2014.

\bibitem{eco}
C.~Beckel, W.~Kleiminger, R.~Cicchetti, T.~Staake, and S.~Santini.
\newblock The eco data set and the performance of non-intrusive load monitoring
  algorithms.
\newblock In {\em Proceedings of the First ACM International Conference on
  Embedded Systems For Energy-Efficient Buildings}. ACM, 2014.

\bibitem{thermosense}
A.~Beltran, V.~L. Erickson, and A.~E. Cerpa.
\newblock Thermosense: Occupancy thermal based sensing for hvac control.
\newblock In {\em Proceedings of the 5th ACM Workshop on Embedded Systems For
  Energy-Efficient Buildings}, pages 1--8. ACM, 2013.

\bibitem{multimodal}
M.~Berg'es and A.~Rowe.
\newblock Appliance classification and energy management using multi-modal
  sensing.
\newblock In {\em Proceedings of the Third ACM Workshop on Embedded Sensing
  Systems for Energy-Efficiency in Buildings}, BuildSys '11, pages 51--52, New
  York, NY, USA, 2011. ACM.

\bibitem{holy_grail}
K.~Carrie~Armel, A.~Gupta, G.~Shrimali, and A.~Albert.
\newblock Is disaggregation the holy grail of energy efficiency? the case of
  electricity.
\newblock {\em Energy Policy}, 52:213--234, 2013.

\bibitem{bidgely_2013}
P.~Chakravarty and A.~Gupta.
\newblock Impact of energy disaggregation on consumer behavior.
\newblock 2013.

\bibitem{niom_umass}
D.~Chen, S.~Barker, A.~Subbaswamy, D.~Irwin, and P.~Shenoy.
\newblock Non-intrusive occupancy monitoring using smart meters.
\newblock In {\em Proceedings of the 5th ACM Workshop on Embedded Systems For
  Energy-Efficient Buildings}, pages 1--8. ACM, 2013.

\bibitem{smart_grid_data_deluge}
B.~Ciara.
\newblock The smart grid data deluge, June 2011.

\bibitem{darby_2006}
S.~Darby.
\newblock The effectiveness of feedback on energy consumption.
\newblock {\em A Review for DEFRA of the Literature on Metering, Billing and
  direct Displays}, 2006.

\bibitem{boss}
S.~Dawson-Haggerty, A.~Krioukov, J.~Taneja, S.~Karandikar, G.~Fierro,
  N.~Kitaev, and D.~Culler.
\newblock Boss: building operating system services.

\bibitem{scale}
S.~Dawson-Haggerty, S.~Lanzisera, J.~Taneja, R.~Brown, and D.~Culler.
\newblock @ scale: Insights from a large, long-lived appliance energy wsn.
\newblock In {\em Proceedings of the 11th international conference on
  Information Processing in Sensor Networks}, pages 37--48. ACM, 2012.

\bibitem{lightwise}
D.~T. Delaney, G.~M. O'Hare, and A.~G. Ruzzelli.
\newblock Evaluation of energy-efficiency in lighting systems using sensor
  networks.
\newblock In {\em Proceedings of the First ACM Workshop on Embedded Sensing
  Systems for Energy-Efficiency in Buildings}, pages 61--66. ACM, 2009.

\bibitem{building_automation_1}
D.~Dietrich, D.~Bruckner, G.~Zucker, and P.~Palensky.
\newblock Communication and computation in buildings: A short introduction and
  overview.
\newblock {\em Industrial Electronics, IEEE Transactions on},
  57(11):3577--3584, 2010.

\bibitem{reduce}
F.~Englert, I.~Diaconita, A.~Reinhardt, A.~Alhamoud, R.~Meister, L.~Backert,
  and R.~Steinmetz.
\newblock Reduce the number of sensors: Sensing acoustic emissions to estimate
  appliance energy usage.
\newblock In {\em Proceedings of the 5th ACM Workshop on Embedded Systems For
  Energy-Efficient Buildings}, pages 1--8. ACM, 2013.

\bibitem{thermovote}
V.~L. Erickson and A.~E. Cerpa.
\newblock Thermovote: Participatory sensing for efficient building hvac
  conditioning.
\newblock 2012.

\bibitem{country_india}
M.~Evans, B.~Shui, and S.~Somasundaram.
\newblock {\em Country report on building energy codes in india}.
\newblock Pacific Northwest National Laboratory, 2009.

\bibitem{country_korea}
M.~Evans, H.~C. UMD, B.~Shui, and S.~Lee.
\newblock Country report on building energy codes in republic of korea.
\newblock {\em Pacific Northwest National Laboratory}, 2009.

\bibitem{socket}
T.~Ganu, D.~Rahayu, D.~P. Seetharam, R.~Kunnath, A.~P. Kumar, V.~Arya, S.~A.
  Husain, and S.~Kalyanaraman.
\newblock Socketwatch: An autonomous appliance monitoring system.

\bibitem{fhmm}
Z.~Ghahramani and M.~I. Jordan.
\newblock Factorial hidden markov models.
\newblock {\em Machine learning}, 29(2-3):245--273, 1997.

\bibitem{picogrid}
S.~K. Ghai, Z.~Charbiwala, S.~Mylavarapu, D.~P. Seetharamakrishnan, and
  R.~Kunnath.
\newblock Dc picogrids: a case for local energy storage for uninterrupted power
  to dc appliances.
\newblock In {\em Proceedings of the fourth international conference on Future
  energy systems}, pages 27--38. ACM, 2013.

\bibitem{home_automation_survey}
C.~Gomez and J.~Paradells.
\newblock Wireless home automation networks: A survey of architectures and
  technologies.
\newblock {\em IEEE Communications Magazine}, 48(6):92--101, 2010.

\bibitem{electrisense}
S.~Gupta, M.~S. Reynolds, and S.~N. Patel.
\newblock Electrisense: single-point sensing using emi for electrical event
  detection and classification in the home.
\newblock In {\em Proceedings of the 12th ACM international conference on
  Ubiquitous computing}, pages 139--148. ACM, 2010.

\bibitem{country_us}
M.~A. Halverson, B.~Shui, and M.~Evans.
\newblock Country report on building energy codes in the united states.
\newblock {\em Pacific Northwest National Laboratory}, 2009.

\bibitem{isleep}
T.~Hao, G.~Xing, and G.~Zhou.
\newblock isleep: unobtrusive sleep quality monitoring using smartphones.
\newblock In {\em Proceedings of the 11th ACM Conference on Embedded Networked
  Sensor Systems}, page~4. ACM, 2013.

\bibitem{hart_1992}
G.~W. Hart.
\newblock Nonintrusive appliance load monitoring.
\newblock {\em Proceedings of the IEEE}, 80(12):1870--1891, 1992.

\bibitem{apportionment}
S.~Hay and A.~Rice.
\newblock The case for apportionment.
\newblock In {\em Proceedings of the First ACM Workshop on Embedded Sensing
  Systems for Energy-Efficiency in Buildings}, pages 13--18. ACM, 2009.

\bibitem{hitchhikers_residential}
T.~W. Hnat, V.~Srinivasan, J.~Lu, T.~I. Sookoor, R.~Dawson, J.~Stankovic, and
  K.~Whitehouse.
\newblock The hitchhiker's guide to successful residential sensing deployments.
\newblock In {\em Proceedings of the 9th ACM Conference on Embedded Networked
  Sensor Systems}, pages 232--245. ACM, 2011.

\bibitem{pecan}
C.~Holcomb.
\newblock {Pecan Street Inc.: A Test-bed for NILM}.
\newblock In {\em International Workshop on Non-Intrusive Load Monitoring},
  Pittsburgh, PA, USA, 2012.

\bibitem{spatial}
D.~Hong, J.~Ortiz, K.~Whitehouse, and D.~Culler.
\newblock Towards automatic spatial verification of sensor placement in
  buildings.
\newblock 2013.

\bibitem{hbci}
J.~Hsu, P.~Mohan, X.~Jiang, J.~Ortiz, S.~Shankar, S.~Dawson-Haggerty, and
  D.~Culler.
\newblock Hbci: human-building-computer interaction.
\newblock In {\em Proceedings of the 2nd ACM Workshop on Embedded Sensing
  Systems for Energy-Efficiency in Building}, pages 55--60. ACM, 2010.

\bibitem{insteon_ha}
D.~Irwin, S.~Barker, A.~Mishra, P.~Shenoy, A.~Wu, and J.~Albrecht.
\newblock Exploiting home automation protocols for load monitoring in smart
  buildings.
\newblock In {\em Proceedings of the Third ACM Workshop on Embedded Sensing
  Systems for Energy-Efficiency in Buildings}, BuildSys '11, pages 7--12, New
  York, NY, USA, 2011. ACM.

\bibitem{johnson_2013}
M.~J. Johnson and A.~S. Willsky.
\newblock {Bayesian Nonparametric Hidden Semi-Markov Models}.
\newblock {\em Journal of Machine Learning Research}, 14:673--701, 2013.

\bibitem{body_1}
E.~Jovanov, A.~Milenkovic, C.~Otto, and P.~C. De~Groen.
\newblock A wireless body area network of intelligent motion sensors for
  computer assisted physical rehabilitation.
\newblock {\em Journal of NeuroEngineering and rehabilitation}, 2(1):6, 2005.

\bibitem{energy_water}
J.~Kadengal, S.~Thirunavukkarasu, A.~Vasan, V.~Sarangan, and
  A.~Sivasubramaniam.
\newblock The energy-water nexus in campuses.
\newblock In {\em Proceedings of the 5th ACM Workshop on Embedded Systems For
  Energy-Efficient Buildings}, BuildSys'13, pages 15:1--15:8, New York, NY,
  USA, 2013. ACM.

\bibitem{energy_auditing}
A.~Kamthe, V.~Erickson, M.~A. Carreira-Perpin{\'a}n, and A.~Cerpa.
\newblock Enabling building energy auditing using adapted occupancy models.
\newblock In {\em Proceedings of the Third ACM Workshop on Embedded Sensing
  Systems for Energy-Efficiency in Buildings}, pages 31--36. ACM, 2011.

\bibitem{building_automation}
W.~Kastner, G.~Neugschwandtner, S.~Soucek, and H.~Newmann.
\newblock Communication systems for building automation and control.
\newblock {\em Proceedings of the IEEE}, 93(6):1178--1203, 2005.

\bibitem{Kavousian2013a}
A.~Kavousian, R.~Rajagopal, and M.~Fischer.
\newblock {Determinants of residential electricity consumption: Using smart
  meter data to examine the effect of climate, building characteristics,
  appliance stock, and occupants' behavior}.
\newblock {\em Energy}, 55(0):184 -- 194, 2013.

\bibitem{ukpd}
J.~Kelly and W.~Knottenbelt.
\newblock Uk-dale: A dataset recording uk domestic appliance-level electricity
  demand and whole-house demand.
\newblock {\em arXiv preprint arXiv:1404.0284}, 2014.

\bibitem{granger}
Y.~Kim, R.~Balani, H.~Zhao, and M.~B. Srivastava.
\newblock Granger causality analysis on ip traffic and circuit-level energy
  monitoring.
\newblock In {\em Proceedings of the 2nd ACM Workshop on Embedded Sensing
  Systems for Energy-Efficiency in Building}, pages 43--48. ACM, 2010.

\bibitem{viridiscope}
Y.~Kim, T.~Schmid, Z.~M. Charbiwala, and M.~B. Srivastava.
\newblock Viridiscope: design and implementation of a fine grained power
  monitoring system for homes.
\newblock In {\em Proceedings of the 11th international conference on
  Ubiquitous computing}, pages 245--254. ACM, 2009.

\bibitem{challenges_residential}
Y.~Kim, T.~Schmid, M.~B. Srivastava, and Y.~Wang.
\newblock Challenges in resource monitoring for residential spaces.
\newblock In {\em Proceedings of the First ACM Workshop on Embedded Sensing
  Systems for Energy-Efficiency in Buildings}, pages 1--6. ACM, 2009.

\bibitem{niom_eth}
W.~Kleiminger, C.~Beckel, T.~Staake, and S.~Santini.
\newblock Occupancy detection from electricity consumption data.
\newblock In {\em Proceedings of the 5th ACM Workshop on Embedded Systems For
  Energy-Efficient Buildings}, pages 1--8. ACM, 2013.

\bibitem{kolter_2012}
J.~Z. Kolter and T.~Jaakkola.
\newblock {Approximate Inference in Additive Factorial HMMs with Application to
  Energy Disaggregation}.
\newblock In {\em Proceedings of the International Conference on Artificial
  Intelligence and Statistics}, pages 1472--1482, La Palma, Canary Islands,
  2012.

\bibitem{redd}
J.~Z. Kolter and M.~J. Johnson.
\newblock {REDD: A public data set for energy disaggregation research}.
\newblock In {\em Proceedings of 1st KDD Workshop on Data Mining Applications
  in Sustainability}, San Diego, CA, USA, 2011.

\bibitem{near_optimal}
A.~Krause, A.~Singh, and C.~Guestrin.
\newblock Near-optimal sensor placements in gaussian processes: Theory,
  efficient algorithms and empirical studies.
\newblock {\em The Journal of Machine Learning Research}, 9:235--284, 2008.

\bibitem{lighting}
A.~Krioukov, S.~Dawson-Haggerty, L.~Lee, O.~Rehmane, and D.~Culler.
\newblock A living laboratory study in personalized automated lighting
  controls.
\newblock In {\em Proceedings of the Third ACM Workshop on Embedded Sensing
  Systems for Energy-Efficiency in Buildings}, pages 1--6. ACM, 2011.

\bibitem{bas}
A.~Krioukov, G.~Fierro, N.~Kitaev, and D.~Culler.
\newblock Building application stack (bas).
\newblock In {\em Proceedings of the Fourth ACM Workshop on Embedded Sensing
  Systems for Energy-Efficiency in Buildings}, pages 72--79. ACM, 2012.

\bibitem{coughsense}
E.~C. Larson, T.~Lee, S.~Liu, M.~Rosenfeld, and S.~N. Patel.
\newblock Accurate and privacy preserving cough sensing using a low-cost
  microphone.
\newblock In {\em Proceedings of the 13th international conference on
  Ubiquitous computing}, pages 375--384. ACM, 2011.

\bibitem{mnist}
Y.~LeCun and C.~Cortes.
\newblock Mnist handwritten digit database.
\newblock {\em AT\&T Labs [Online]. Available: http://yann. lecun.
  com/exdb/mnist}, 2010.

\bibitem{smart_thermostat}
J.~Lu, T.~Sookoor, V.~Srinivasan, G.~Gao, B.~Holben, J.~Stankovic, E.~Field,
  and K.~Whitehouse.
\newblock The smart thermostat: using occupancy sensors to save energy in
  homes.
\newblock In {\em Proceedings of the 8th ACM Conference on Embedded Networked
  Sensor Systems}, pages 211--224. ACM, 2010.

\bibitem{berds}
M.~Maasoumy, B.~M. Sanandaji, K.~Poolla, and A.~S. Vincentelli.
\newblock Berds-berkeley energy disaggregation data set.

\bibitem{ampds}
S.~Makonin, F.~Popowich, L.~Bartram, B.~Gill, and I.~V. Bajic.
\newblock {AMPds: A Public Dataset for Load Disaggregation and Eco-Feedback
  Research}.
\newblock In {\em IEEE Electrical Power and Energy Conference}, Halifax, NS,
  Canada, 2013.

\bibitem{leap}
D.~McIntire, K.~Ho, B.~Yip, A.~Singh, W.~Wu, and W.~J. Kaiser.
\newblock The low power energy aware processing (leap) embedded networked
  sensor system.
\newblock In {\em Proceedings of the 5th international conference on
  Information processing in sensor networks}, pages 449--457. ACM, 2006.

\bibitem{private_memoirs}
A.~Molina-Markham, P.~Shenoy, K.~Fu, E.~Cecchet, and D.~Irwin.
\newblock Private memoirs of a smart meter.
\newblock In {\em Proceedings of the 2nd ACM workshop on embedded sensing
  systems for energy-efficiency in building}, pages 61--66. ACM, 2010.

\bibitem{greend}
A.~Monacchi, D.~Egarter, W.~Elmenreich, S.~D'Alessandro, and A.~M. Tonello.
\newblock Greend: An energy consumption dataset of households in italy and
  austria.
\newblock {\em arXiv preprint arXiv:1405.3100}, 2014.

\bibitem{mleplus}
T.~X. Nghiem.
\newblock {MLE+}: a {Matlab}-{EnergyPlus} co-simulation interface.
\newblock \url{http://www.seas.upenn.edu/~nghiem/mleplus.html}.

\bibitem{energy_buildings_survey}
T.~A. Nguyen and M.~Aiello.
\newblock Energy intelligent buildings based on user activity: A survey.
\newblock {\em Energy and buildings}, 56:244--257, 2013.

\bibitem{indoor_2}
V.~Otsason, A.~Varshavsky, A.~LaMarca, and E.~De~Lara.
\newblock Accurate gsm indoor localization.
\newblock In {\em UbiComp 2005: Ubiquitous Computing}, pages 141--158.
  Springer, 2005.

\bibitem{body_2}
C.~Otto, A.~Milenkovic, C.~Sanders, and E.~Jovanov.
\newblock System architecture of a wireless body area sensor network for
  ubiquitous health monitoring.
\newblock {\em Journal of Mobile Multimedia}, 1(4):307--326, 2006.

\bibitem{conference}
K.~Padmanabh, A.~Malikarjuna, V, S.~Sen, S.~P. Katru, A.~Kumar, S.~P. C, S.~K.
  Vuppala, and S.~Paul.
\newblock isense: A wireless sensor network based conference room management
  system.
\newblock In {\em Proceedings of the First ACM Workshop on Embedded Sensing
  Systems for Energy-Efficiency in Buildings}, BuildSys '09, pages 37--42, New
  York, NY, USA, 2009. ACM.

\bibitem{parson_2012}
O.~Parson, S.~Ghosh, M.~Weal, and A.~Rogers.
\newblock {Non-intrusive load monitoring using prior models of general
  appliance types}.
\newblock In {\em Proceedings of the 26th AAAI Conference on Artificial
  Intelligence}, pages 356--362, Toronto, ON, Canada, 2012.

\bibitem{sust}
L.~Pereira, F.~Quintal, R.~Gon{\c{c}}alves, and N.~J. Nunes.
\newblock Sustdata: A public dataset for ict4s electric energy research.
\newblock In {\em ICT for Sustainability 2014 (ICT4S-14)}. Atlantis Press,
  2014.

\bibitem{novel}
J.~Pierce and E.~Paulos.
\newblock Beyond energy monitors: Interaction, energy, and emerging energy
  systems.
\newblock In {\em Proceedings of the SIGCHI Conference on Human Factors in
  Computing Systems}, CHI '12, pages 665--674, New York, NY, USA, 2012. ACM.

\bibitem{building_automation_protocols}
J.~Piper.
\newblock Bacnet, lonmark and modbus: How and why they work.
\newblock
  \url{http://facilitiesnet.com/buildingautomation/article/BACnet-LonMark-and-Modbus-How-and-Why-They-Work--7712}.
\newblock Accessed: 2014-03-25.

\bibitem{hot_water_dj}
M.~A. Prodhan and K.~Whitehouse.
\newblock Hot water dj: Saving energy by pre-mixing hot water.
\newblock In {\em Proceedings of the Fourth ACM Workshop on Embedded Sensing
  Systems for Energy-Efficiency in Buildings}, pages 91--98. ACM, 2012.

\bibitem{rahayu_2012}
D.~Rahayu, B.~Narayanaswamy, S.~Krishnaswamy, C.~Labbe, and D.~P. Seetharam.
\newblock {Learning to be energy-wise: Discriminative methods for load
  disaggregation}.
\newblock In {\em 3rd International Conference on Future Energy Systems}, pages
  1--4, 2012.

\bibitem{tracebase}
A.~Reinhardt, P.~Baumann, D.~Burgstahler, M.~Hollick, H.~Chonov, M.~Werner, and
  R.~Steinmetz.
\newblock {On the Accuracy of Appliance Identification Based on Distributed
  Load Metering Data}.
\newblock In {\em Proceedings of the 2nd IFIP Conference on Sustainable
  Internet and ICT for Sustainability (SustainIT)}, pages 1--9, 2012.

\bibitem{RicharddeDear2002}
{Richard de Dear} and {Melissa Hart}.
\newblock {Appliance Electricity End-Use: Weather and Climate Sensitivity}.
\newblock Technical report, Sustainable Energy Group, Australian Greenhouse
  Office, 2002.

\bibitem{myjuolo}
A.~Rogers, R.~Wilcock, S.~Ghosh, and N.~R. Jennings.
\newblock A scalable low-cost solution to provide personalized home heating
  advice to households.
\newblock In {\em Proceedings of the Fourth ACM Workshop on Embedded Sensing
  Systems for Energy-Efficiency in Buildings}, pages 211--212. ACM, 2013.

\bibitem{energylens}
M.~Saha, S.~Thakur, A.~Singh, and Y.~Agarwal.
\newblock {EnergyLens: Combining Smartphones with Electricity Meter for
  Accurate Activity Detection and User Annotation}.
\newblock In {\em Fifth International Conference on Future Energy Systems
  (e-Energy)}, pages 289--300. ACM, 2014.

\bibitem{satyanarayanan}
M.~Satyanarayanan.
\newblock Pervasive computing: Vision and challenges.
\newblock {\em Personal Communications, IEEE}, 8(4):10--17, 2001.

\bibitem{copolan}
A.~Schoofs, D.~T. Delaney, G.~M. O'Hare, and A.~G. Ruzzelli.
\newblock Copolan: non-invasive occupancy profiling for preliminary assessment
  of hvac fixed timing strategies.
\newblock In {\em Proceedings of the Third ACM Workshop on Embedded Sensing
  Systems for Energy-Efficiency in Buildings}, pages 25--30. ACM, 2011.

\bibitem{annot}
A.~Schoofs, A.~Guerrieri, D.~T. Delaney, G.~O'Hare, and A.~G. Ruzzelli.
\newblock Annot: Automated electricity data annotation using wireless sensor
  networks.
\newblock In {\em Sensor Mesh and Ad Hoc Communications and Networks (SECON),
  2010 7th Annual IEEE Communications Society Conference on}, pages 1--9. IEEE,
  2010.

\bibitem{preheat}
J.~Scott, A.~Bernheim~Brush, J.~Krumm, B.~Meyers, M.~Hazas, S.~Hodges, and
  N.~Villar.
\newblock Preheat: controlling home heating using occupancy prediction.
\newblock In {\em Proceedings of the 13th international conference on
  Ubiquitous computing}, pages 281--290. ACM, 2011.

\bibitem{country_china}
B.~Shui, M.~Evans, H.~Lin, W.~Jiang, B.~Liu, B.~Song, and S.~Somasundaram.
\newblock {\em Country Report on Building Energy Codes in China}.
\newblock Pacific Northwest National Laboratory, 2009.

\bibitem{country_australia}
B.~Shui, M.~Evans, and S.~Somasundaram.
\newblock {\em Country report on building energy codes in Australia}.
\newblock Pacific Northwest National Laboratory, 2009.

\bibitem{watersense}
V.~Srinivasan, J.~Stankovic, and K.~Whitehouse.
\newblock Watersense: Water flow disaggregation using motion sensors.
\newblock In {\em Proceedings of the Third ACM Workshop on Embedded Sensing
  Systems for Energy-Efficiency in Buildings}, pages 19--24. ACM, 2011.

\bibitem{fixturefinder}
V.~Srinivasan, J.~Stankovic, and K.~Whitehouse.
\newblock Fixturefinder: Discovering the existence of electrical and water
  fixtures.
\newblock In {\em Proceedings of the 12th international conference on
  Information processing in sensor networks}, pages 115--128. ACM, 2013.

\bibitem{indoor_1}
M.~Sugano, T.~Kawazoe, Y.~Ohta, and M.~Murata.
\newblock Indoor localization system using rssi measurement of wireless sensor
  network based on zigbee standard.
\newblock {\em Target}, 538:050, 2006.

\bibitem{softgreen}
L.~V. Thanayankizil, S.~K. Ghai, D.~Chakraborty, and D.~P. Seetharam.
\newblock Softgreen: Towards energy management of green office buildings with
  soft sensors.
\newblock In {\em Communication Systems and Networks (COMSNETS), 2012 Fourth
  International Conference on}, pages 1--6. IEEE, 2012.

\bibitem{plug_load}
T.~Weng, B.~Balaji, S.~Dutta, R.~Gupta, and Y.~Agarwal.
\newblock Managing plug-loads for demand response within buildings.
\newblock In {\em Proceedings of the Third ACM Workshop on Embedded Sensing
  Systems for Energy-Efficiency in Buildings}, pages 13--18. ACM, 2011.

\bibitem{building_depot2}
T.~Weng, A.~Nwokafor, and Y.~Agarwal.
\newblock Buildingdepot 2.0: An integrated management system for building
  analysis and control.
\newblock In {\em Proceedings of the 5th ACM Workshop on Embedded Systems For
  Energy-Efficient Buildings}, pages 1--8. ACM, 2013.

\bibitem{zeifman_2011}
M.~Zeifman and K.~Roth.
\newblock {Nonintrusive appliance load monitoring: Review and outlook}.
\newblock {\em IEEE Transactions on Consumer Electronics}, 57(1):76--84, 2011.

\bibitem{hes}
J.-P. Zimmermann, M.~Evans, J.~Griggs, N.~King, L.~Harding, P.~Roberts, and
  C.~Evans.
\newblock {Household Electricity Survey. A study of domestic electrical product
  usage}.
\newblock Technical Report R66141, DEFRA, May 2012.

\bibitem{survey_nilm}
A.~Zoha, A.~Gluhak, M.~A. Imran, and S.~Rajasegarar.
\newblock Non-intrusive load monitoring approaches for disaggregated energy
  sensing: A survey.
\newblock {\em Sensors}, 12(12):16838--16866, 2012.

\end{thebibliography}
}

\clearpage

% % Commenetd the appendix for now % %
%\appendix
%\section{Planned work}
%\noindent In this section we briefly present the future work we intend to carry filling in some of the gaps mentioned above. We have spent considerable efforts towards a home deployment in New Delhi. As a result, we have been able to quantify several problems such as unreliable grid and network. In our previous work, we had presented initial analysis on detecting power outages. We wish to extend that study across more homes across different zones in the city. Further, we wish to take this research forward by modeling the factors which can help predict power outages. The proposed approach needs to be developed in a cost-effective fashion owing to the intended scale of deployment. We also plan to take this a step beyond and finding out ways to mitigate such power outages at a society or home cluster level. 

%On the NILM front we wish to carry a user study understanding the benefits which disaggregated real time information can provide. To this end, we plan to leverage the faculty apartment already have deployed smart meters across several faculty apartments in our campus and have developed necessary software for performing this disaggregation.

\end{document}